\documentclass[namedreferences,hyperref,optionalrh,solaromanenum,linenumbers]{spr-sola}
\usepackage{amsmath,amssymb}
\usepackage{multirow}
\usepackage{lipsum}
\usepackage{placeins}
\usepackage{graphicx}
\usepackage{amssymb}        
\usepackage{color}           

\renewcommand{\vec}[1]{{\mathbfit #1}}

\chardef\us=`\_

\begin{document}

\begin{article}
\begin{opening}

\title{Observation and Modeling of Small Spatial Structures of Solar Radio Noise Storms using the uGMRT}

\author[addressref={aff1},corref,email={surajit.mondal@njit.edu}]{\inits{Surajit }\fnm{Surajit }\snm{Mondal}\orcid{0000-0002-2325-5298}}
\author[addressref={aff2,aff3},corref]{\inits{P. Z. }\fnm{Peijin }\snm{Zhang} \orcid{0000-0001-6855-5799}}
\author[addressref={aff2,aff4},corref]{\inits{D. K. }\fnm{Devojyoti }\snm{Kansabanik}\orcid{0000-0001-8801-9635}}
\author[addressref={aff5},corref]{\inits{D. O. }\fnm{Divya }\snm{Oberoi}\orcid{0000-0002-4768-9058}}
\author[addressref={aff6},corref]{\inits{G. P. }\fnm{Gillian }\snm{Pearce}\orcid{0000-0003-1428-2021}}

\address[id=aff1]{Center for Solar-Terrestrial Research, New Jersey Institute of Technology, 323 M L King Jr Boulevard, Newark, NJ 07102-1982, USA}

\address[id=aff2]{Cooperative Programs for the Advancement of Earth System Science, University Corporation for Atmospheric Research, 3090 Center Green Dr, Boulder, CO, USA 80301}

\address[id=aff3]{NASA Jack Eddy fellow hosted at the Center for Solar-Terrestrial Research, New Jersey Institute of Technology, 323 M L King Jr Boulevard, Newark, NJ 07102-1982, USA}

\address[id=aff4]{NASA Jack Eddy fellow hosted at the Johns Hopkins University Applied Physics Laboratory, 11001 Johns Hopkins Rd, Laurel, USA 20723}

\address[id=aff5]{National Centre for Radio Astrophysics, Tata Institute of Fundamental Research, S. P. Pune University Campus, Pune 411007, India}

\address[id=aff6]{Department of Mechatronics and Biomedical Engineering, College of Engineering and Physical Sciences, Aston University, Birmingham, B4 7ET}

\runningauthor{Mondal et al.}
\runningtitle{Observation and Modeling of Small Spatial Structures of Solar Radio Noise Storms}

\begin{abstract}
One of the most commonly observed solar radio sources in the metric and decametric wavelengths is the solar noise storm. These are generally associated with active regions and are believed to be powered by the plasma emission mechanism. Since plasma emission emits primarily at the fundamental and harmonic of the local plasma frequency, it is significantly affected by density inhomogeneities in the solar corona. The source can become significantly scatter-broadened due to the multi-path propagation caused by refraction from the density inhomogeneities. Past observational and theoretical estimates suggest some minimum observable source size in the solar corona. The details of this limit, however, depends on the modeling approach and details of the coronal turbulence model chosen. Hence pushing the minimum observable source size to smaller values can help constrain the plasma environment of the observed sources. In this work, we for the first time, use data from the upgraded Giant Metrewave Radio Telescope in the 250--500 MHz band, to determine multiple instances of very small-scale structures in the noise storms. We also find that these structures are stable over timescales of 15--30 minutes. By comparing the past observations of Type III radio bursts and noise storms, we hypothesize that the primary reason behind the detection of these small sources in noise storm is due to the local environment of the noise storm. We also build an illustrative model and propose some conditions under which the minimum observable source size predicted by theoretical models, can be lowered significantly.
\end{abstract}

\end{opening}

\section{Introduction}\label{sec:introduction} 
Noise storms are one of the most commonly observed solar radio sources in the metric and decametric wavelengths. These are generally associated with active regions \citep[e.g.][]{mugundhan2018, mccauley2019, mohan2019b} and often last for several days \citep{elgaroy1977}. Noise storms are characterized by short duration ($\sim 0.1–10$ s) intense narrowband ($\sim$ a few MHz) bursts (type--I bursts) superposed on a long duration wideband ($\gtrsim 100$ MHz) continuum emission \citep{mclean1967}, and are often strongly circularly polarized \citep[e.g.][]{zlobec1971, ramesh2011, mugundhan2018,mccauley2019}. Noise storms are believed to arise because of plasma emission from nonthermal electrons trapped in coronal loops \citep{sakurai1971, melrose1980}. 

The vast majority of studies of noise storms have  been done using the dynamic spectra and they have been classified as a separate class of solar radio bursts based on their appearance in the time-frequency plane. Imaging studies of noise storms have mostly been done using instruments with low spatial resolution. This is motivated primarily by two reasons -- first, early observations of noise storms, in the early '90s and 2000s, did not reveal the presence of small spatial structures \citep[e.g.][]{lang1987, zlobec1992, kerdraon1988}; and second, the expectation based on theoretical studies is that this emission will be subject to severe scattering, which will smoothen out any small angular scale structures, even if they were to be present \citep[e.g.][]{bastian1994, kontar2019}. 

However, small angular scale structures inside  noise storms have been reported in a few earlier works \citep[e.g.][]{mercier2015,mugundhan2018, mondal2024_noise_storm}. \citet{mondal2024_noise_storm}, henceforth refereed to as M24, observed multiple instances of sources with compact angular sizes, dropping to an angular size as low as $\lesssim 9^"$, in one instance. This is about three times smaller than the smallest reported angular size at similar frequencies prior to M24. Additionally, the minimum observable source size due to scattering is also dependent on the assumptions made and can vary from about an arcsecond to about an arcminute \citep{subramanian2011}. Hence, it is important to understand under what conditions can one observe small scale structures inside noise storms. This is very important for better understanding the turbulent nature of the solar corona, particularly close to the active regions which have more complicated magnetic field and density structures. Additionally, the lack of sufficient angular resolution of the instruments used has always been a limiting factor in associating radio observations at metric wavelengths with high resolution observations at Extreme Ultraviolet (EUV) and X-ray wavelengths. The observed small scale features inside noise storms can throw a new light into the dynamics happening inside noise storms, which when combined with the data from other wavebands, can hopefully provide a more complete picture of the relevant physical processes. 

However, the number of such detections is rather small, and they can easily be regarded as chance alignments in a rather turbulent medium. In fact, M24 pointed out that all the three instances where  small angular scale  structures have been reported in noise storms, were located on/close to the western limb. There is no apparent reason for this. This points to a probable sample bias, which can be only taken care of by having more high angular resolution observations of noise storms, located at different heliocentric longitudes. Only with more such detections, would it become possible to draw statistically significant conclusions about the conditions under which small angular scale structures might be visible in a noise storm. 

With this motivation, we have started a survey of noise storms with the upgraded Giant Metrewave Radio Telescope \citep[uGMRT,][]{Swarup2000, gupta2017} in 250-500 MHz band (band-3).
This band represents an optimal choice between the presence of solar noise storms and the local radio frequency interference (RFI) conditions.
Here we present the results obtained from this pilot survey. Section \ref{sec:cur_stat} discusses the current status of high angular resolution observations of type-I noise storms. The observation and calibration details are provided in Section \ref{sec:obs_cal}. The results obtained from this analysis are provided in Section \ref{sec:imaging_results}. In Section \ref{sec:discussion}, we place our results in the overall context of noise storms and the impact of scattering, and finally in Section \ref{sec:conclusion} we present the conclusions from this investigation.

\section{Current Status of High Angular Resolution Imaging of type-I Noise Storms}\label{sec:cur_stat}

\citet{lang1987} used the Karl G. Jansky Very Large Array (VLA) to observe noise storms with an angular resolution of $9^" \times 9^"$, for around five hours duration. They found that the noise storm consisted of multiple sources, with sizes of about $40^"$, located within an extended source. In an observation with even higher angular resolution of $4^"$, no structure smaller than $40^"$ was observed by \citet{zlobec1992} at 333 MHz. The typical source size they reported varied between $40^"-90^"$. \citet{mercier2006} and \citet{mercier2015}, however, observed sources with sizes $31-35^"$ at 236--327 MHz band when they combined data from the legacy Giant Metrewave Radio Telescope \citep[GMRT,][]{Swarup2000} and the Nan\c{c}ay Radio Heliograph\citep[NRH,][]{bonmartin1983,avignon1989,kerdraon1997} to produce snapshot images at a resolution of $20^"$. \citet{mugundhan2018} detected source sizes $\leq 15^"$ at 53 MHz using a two element interferometer with baseline length of $\sim 200$ km. M24 used observations with an angular resolution of $17.5^"\times 9.2^"$ in the frequency range $\sim 217-250$ MHz and demonstrated that for the entire 4 minutes of their observation, there were structures with angular scale $\sim 20^"$. Additionally, they presented one instance, where the measured source size fell to $\lesssim 9^"$.

\section{Observations and Calibration} \label{sec:obs_cal}

\subsection{uGMRT Observation Details}\label{subsec:obs}
The uGMRT is a radio interferometric array consisting of 30 antennas, each 45 meters in diameter, spread over an area $\sim$25 km in diameter located at Khodad, India. The uGMRT provides an angular resolution of a few arcseconds over its entire observing frequency range. We observed the Sun using the band-3 receivers of the uGMRT covering the frequency range 250--500 MHz. These data were taken under the project code DDTC358. 

Sensitive radio telescopes are optimized to observe astronomical sources. Since the Sun has orders of magnitude higher flux density compared to the typical astronomical radio sources, one needs to attenuate the solar signal to maintain the entire signal chain of the instrument in the linear regime. Based on several tests done a-priori for developing a standard observation procedure of solar observations using the uGMRT (Kansabanik et al., in prep), we used an attenuation of 30 dB for observing in the uGMRT band-3. 
During calibrator observations, we have used uGMRT online RFI filtering system \citep{Buch2022,Buch2023}. However, since the solar emission can mimic RFI on spectro-temporal domain, online RFI filtering is switched of during the solar scans.

To select the observations to be analysed, we examined the NRH database\footnote{\url{https://secchirh.obspm.fr/spip.php?article11}} to identify the days on which noise storms were observed. 
All days on which noise storms were seen in NRH data were analysed. The corresponding observing details are summarized in Table \ref{tab:obs_details}.

\begin{table}[]
    \centering
    \begin{tabular}{|p{1.8cm}|p{1.2cm}|p{1.2cm}|p{1.4cm}|p{1.4cm}|p{1cm}|}
    \hline
         Date & Start time (UTC) & End time (UTC)& Integration  time (s)  & Frequency  resolution (kHz)  & Smallest scale \\
              &  &  &             &               & \\
         \hline \hline
         June 4, 2024 & 08:59:25 & 09:42:27 & 2.68 & 97.656 &  $11^"$\\
         June 7, 2024 & 09:51:37 & 11:41:40 & 2.68 & 195.312 & $\sim 8^"$\\
         June 11, 2024 & 09:41:5 & 10:55:13 & 2.68 & 195.312 & $15^"$\\
         \hline
    \end{tabular}
    \caption{Some key details of the observations are presented here. Smallest size for June 7 is indicative only, and more details are provided in Section \ref{subsec:june7}.}
    \label{tab:obs_details}
\end{table}

\subsection{Calibration procedure} \label{subsec:calibration}
\begin{figure}
    \centering
    \includegraphics[width=0.7\linewidth]{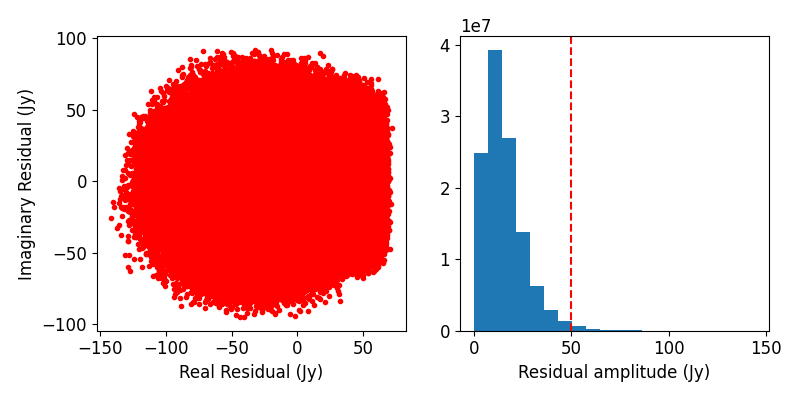}
    \caption{Left panel: The real and imaginary components of the residual visibilities. Right panel: The histogram of the amplitude of the residual visibilities. The dashed line indicates the expected amplitude of the calibrator source.}
    \label{fig:calibration_quality}
\end{figure}

We did the data analysis using the Common Astronomy Software Applications \citep[CASA,][]{CASA2022}. First, we performed basic flagging of bad antennas and bad spectral channels. Next we performed an automated RFI-flagging on the uncalibrated 3C 147 scan using \textsf{tfcrop}\footnote{\href{https://www.aoc.nrao.edu/~rurvashi/TFCrop/TFCropV1/node2.html}{tfcrop algorithm in CASA}} algorithm in \textsf{flagdata} task. Then we determined the frequency-dependent complex gain of the instrument (instrumental bandshape) using the source model of 3C 147 \citep{Perley_2017} and \textsf{bandpass} task. The bandpass solutions were evaluated manually, and the channels which have very low gains were flagged. We corrected the 3C 147 observations for the estimated instrumental bandshape using \textsf{applycal} task. Next we performed an automated flagging on the residual data using \textsf{rflag}\footnote{\href{https://www.aips.nrao.edu/cook.html}{Greisen, Eric, 2011}} algorithm. Next we re-calculated the instrumental bandshape solutions. This process was repeated till we were satisfied with the calibration and no RFI was evident in the residual visibilities. Next we calibrated the 3C 147 scan taken with the same 30 dB attenuation as used for solar scans, following the same procedure. {Figure \ref{fig:calibration_quality} shows the distribution of residual visibility.\footnote{Residual visibility is defined as the difference between the calibrated data and the model data.} The residual visibility is expected to be close to zero, for well calibrated data. In the left panel, we show both the real and imaginary components of the residual visibility, and in the right hand panel the histogram of the amplitudes of the residual visibility. The dashed line in the right panel shows the expected amplitude of the calibrator. The fact that the median residual is close to zero and much smaller than the expected model confirms the high quality of the calibrated data.}
The solutions determined using the 3C 147 observation with the 30 dB attenuator were applied to the solar data. A combination of automated and manual flagging was done to remove RFI. While it is possible that the automated flagging algorithms used may have also discarded some solar bursts, it is not an issue for this work, as we are primarily interested in structures which last for the bulk of the observing duration. 

\section{Imaging the noise storm} \label{sec:imaging_results}

\begin{figure}
    \centering
    \includegraphics[width=0.7\linewidth]{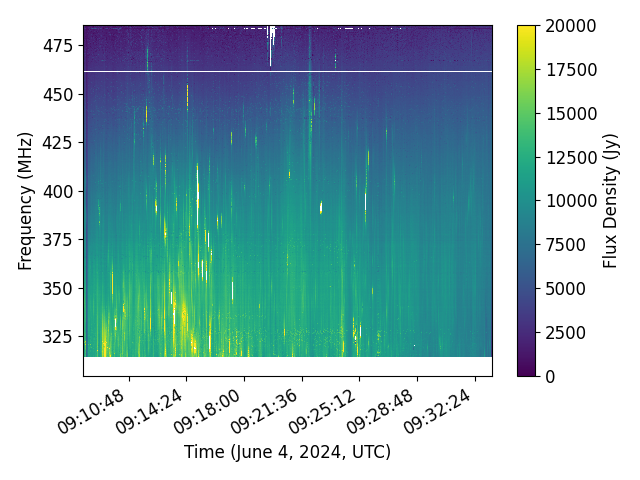}
    \includegraphics[width=0.7\linewidth]{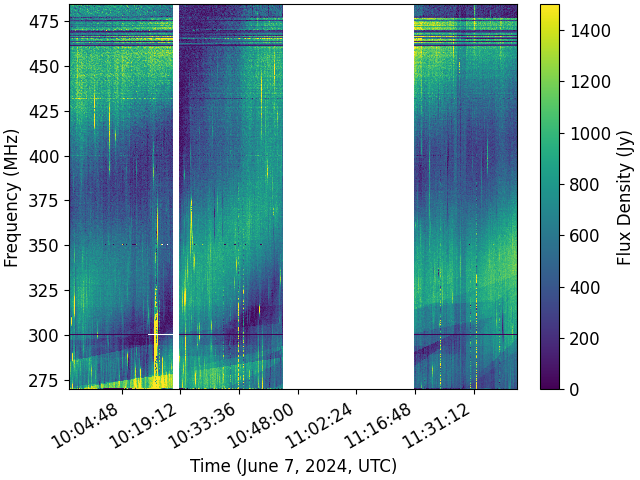}
    \includegraphics[width=0.7\linewidth]{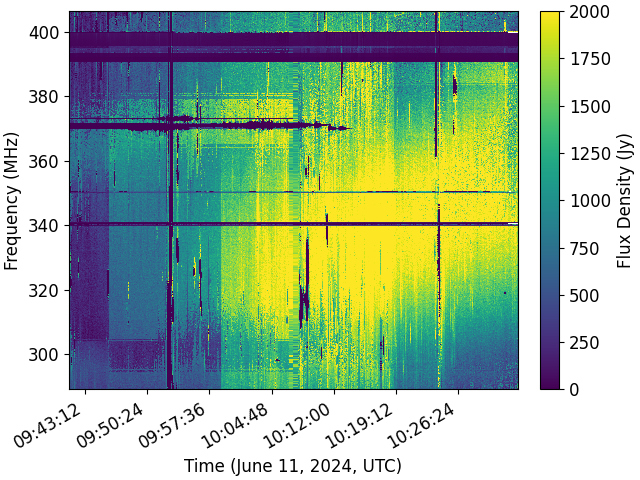}
    \caption{The top, middle and bottom panels show the dynamic spectra corresponding to the observations used here. The dynamic spectra have been generated by averaging data from all baselines with lengths between $300-500\lambda$. }
    \label{fig:dynamic_spectrum}
\end{figure}

{In Figure \ref{fig:dynamic_spectrum} we have presented the dynamic spectra corresponding to the observation times. We have averaged all baselines with lengths between $300-500\lambda$ to produce these dynamic spectra. The total power measurements are often the most affected by RFI and hence are not used. The baselines chosen are sensitive to structures with spatial scales between 6$^{'}$--11$^{'}$, which is much larger than the typical scales of noise storms. Hence these dynamic spectra are expected to provide the total flux density of the noise storms. We also note that that some artefacts are observed in the dynamic spectra corresponding to June 7 and 11. We believe that these artefacts are terrestrial in origin and should be ignored. 

It is evident from the dynamic spectra that there were no strong radio bursts during our observation times. This was also confirmed by manual inspection of the timeseries data for several baselines.} Hence we have averaged over approximately 15--30 minutes to generate a single image. The integration time is determined by the length of an observing scan, after which the phase-centre of the data was changed to a different direction to account for the motion of the Sun in the celestial coordinates. For the weak active emissions seen in our data, the fraction of the noise storm flux density arising from compact structures is rather small. Hence, to facilitate the high-resolution imaging, we first determine the location of the noise storm by making a low-resolution image, using the task \textsf{tclean}. The final high- resolution image is produced by masking the source region as determined from the low-resolution image. { We have also used different baseline ranges and weighting schemes for different images. These play a key role in determining the image fidelity, image sensitivity as well as the image resolution (interested readers are referred to standard textbooks like \citet{thompson2017} for a detailed discussion). Due to the complex interplay between these quantities, the chosen imaging parameters vary across the images presented in this work, and were chosen for each image by trial and error to arrive at the `best' image, which fulfills our objectives. }We provide details of the imaging procedure and results for each day in the following subsections. 

\subsection{June 4, 2024} \label{subsec:june4}
\begin{figure}
    \centering
    \includegraphics[trim={0cm 0cm 3cm 0cm},clip,width=\linewidth]{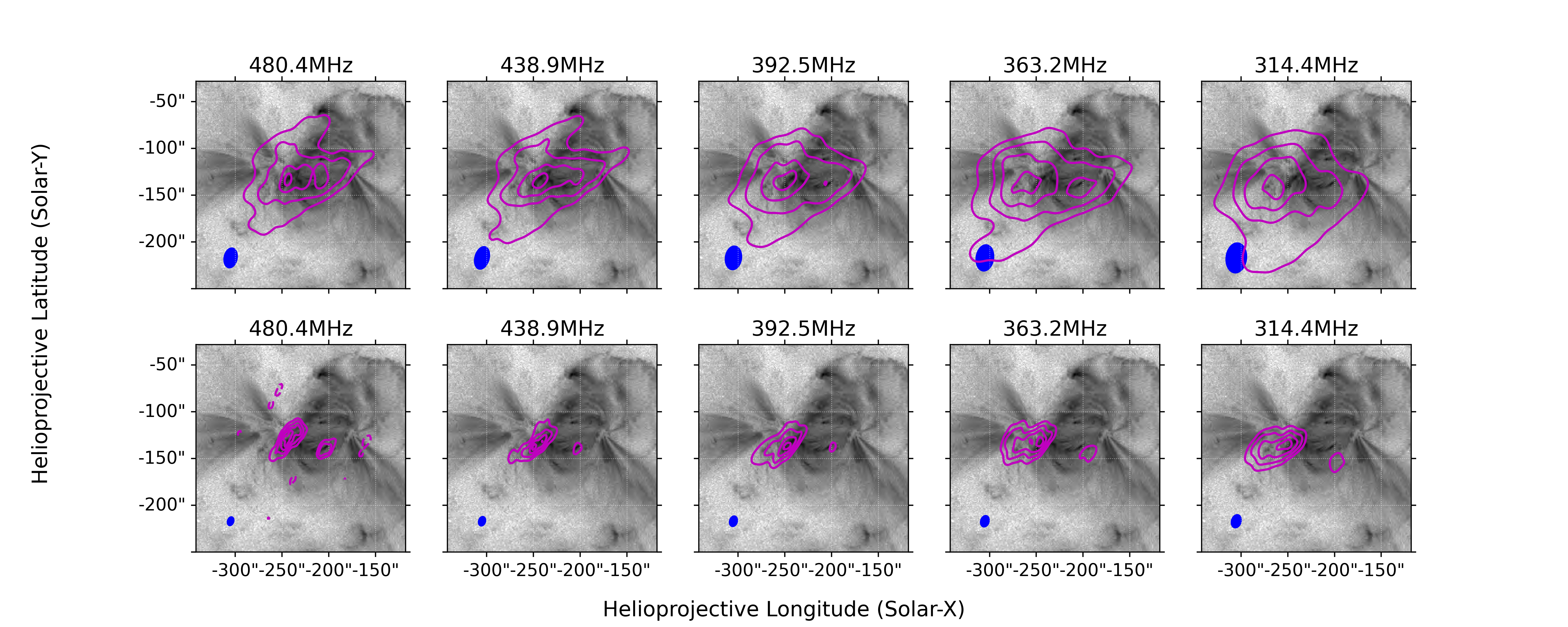}
    \caption{Low (upper panel) and high-resolution (lower panel) images from June 4, 2024. Radio contours at different frequencies have been overlaid on 131\AA$\,$ image. The contour levels are at -0.4, 0.4, 0.6 and 0.8 times the peak flux density. Negative and positive contours are shown with dashed and solid lines. The blue ellipse at the bottom left shows the instrumental resolution of the corresponding image. { The positive contour values increases from the outer contours towards the inner contours. The upper and lower panel images are generated using baselines with lengths between $0.4-10k\lambda$ and $2-20k\lambda$ respectively. \textsf{Uniform} weighting was used in both cases. Approximately 24 MHz and 25 minutes of data were averaged to produce these images.}}
    \label{fig:june4_low_high_res}
\end{figure}

\begin{figure}
    \centering
    \includegraphics[trim={0cm 0cm 2cm 1cm},clip,width=\linewidth]{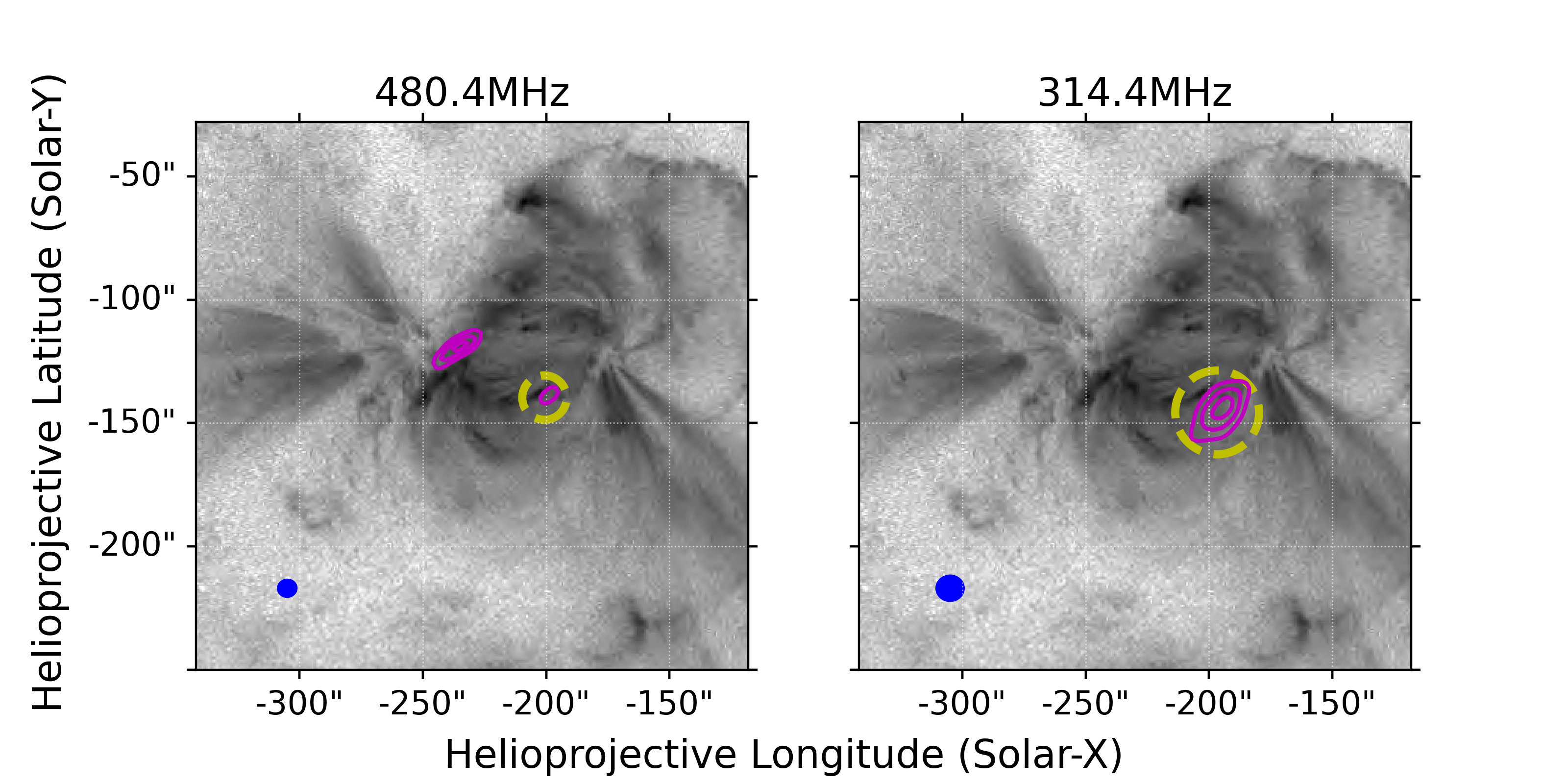}
    \caption{Images from June 4, 2024. High-resolution images of noise storm at the highest (left panel) and the lowest (right panel) observing frequency are shown with contours superposed on nearby 131\AA$\,$ image. The contour levels are at -0.6,0.6,0.8 and 0.95 of the peak in the respective radio image. The blue ellipse at the bottom left shows the instrumental resolution of the corresponding image.  { The positive contour values increases from the outer contours towards the inner contours. Approximately 24 MHz and 25 minutes of data were averaged to produce these images. The central frequency corresponding to each image is indicated in the title of each panel. Images at left and right panels were produced using baselines between $4-25k\lambda$ and $5-20k\lambda$ respectively with \textsf{Briggs} weighting scheme and \textsf{robust} $0.5$.}}
    \label{fig:june4_highest_res}
\end{figure}

The upper and lower panels of Figure \ref{fig:june4_low_high_res} show the low and high-resolution image contours overlaid on a nearby Atmospheric Imaging Assembly \citep[AIA][]{lemen2012} 131$\AA\,$ image. AIA is located on-board the Solar Dynamics Observatory \citep{pesnell2012}. The low-resolution radio images were generated using baselines having lengths between $0.4-10k\lambda$, where $\lambda$ is the observing wavelength. {The high-resolution images were generated using baselines spanning $2-20k\lambda$. Both the low-resolution and high resolution images were generated using \textsf{uniform} weighting.} Approximately 24 MHz and 25 minutes of data were averaged to obtain these images. The contour levels are at -0.4, 0.4, 0.6 and 0.8 times the peak flux density. Negative\footnote{Since sky cannot have negative values, presence of negative values are used as a metric for judging the image fidelity. Presence of negative contour at a particular level generally suggests that the positive values of similar magnitude also have low fidelity.} and positive contours are shown with dashed and solid lines, respectively. { The positive contour values increase from the outer contours towards the inner contours. Absence of any dashed contour indicates the absence of the negative values of the corresponding magnitude.} The radio source has a very similar morphology across the 166 MHz frequency span over the 25 minute averaging interval. Considering the large temporal and spectral span of the radio source, we consider this to be a noise storm. The apparent location of a radio source at these low frequencies can be significantly different from its true location due to ionospheric refraction. We have shifted the radio source to the nearest active region, by giving it a shift of $70^"$ and $10^"$ along helioprojective latitude and longitude, respectively. {The determined shift roughly aligns the extended structure, observed in low-resolution images, with the nearest active region, and is used to correct both the high resolution and low resolution images. We consider this ad-hoc image alignment sufficient for our purpose, as in this work, we have only investigated the morphology of the noise storm sources, which does not require detailed astrometry.} In Figure \ref{fig:june4_highest_res}, we show two images, from the highest and the lowest frequency. Here we have tried to filter out the large-scale structure as much as possible and still detect a compact radio source. The images at 480 and 314 MHz are produced using baselines between $4-25k\lambda$ and $5-20k\lambda$, respectively, with a \textsf{Briggs} \citep{briggs1995} weighting and \textsf{robust} $0.5$. The contour levels are at -0.6, 0.6, 0.8 times the peak flux density in the respective radio images. There is no dashed contour close to the source, attesting to the low level of artefacts in the radio images. The yellow dashed circle shows the region which has been used to fit a Gaussian and determine the deconvolved size of the sources present inside the circle. The deconvolved source sizes obtained for the 480 and 314 MHz images are $20 (5)^" \times 11 (3)^"$ and $30(5)^" \times 16(3)^"$ respectively. 

\begin{figure}
    \centering
    \includegraphics[trim={0cm 0cm 2cm 1.5cm},clip,width=\linewidth]{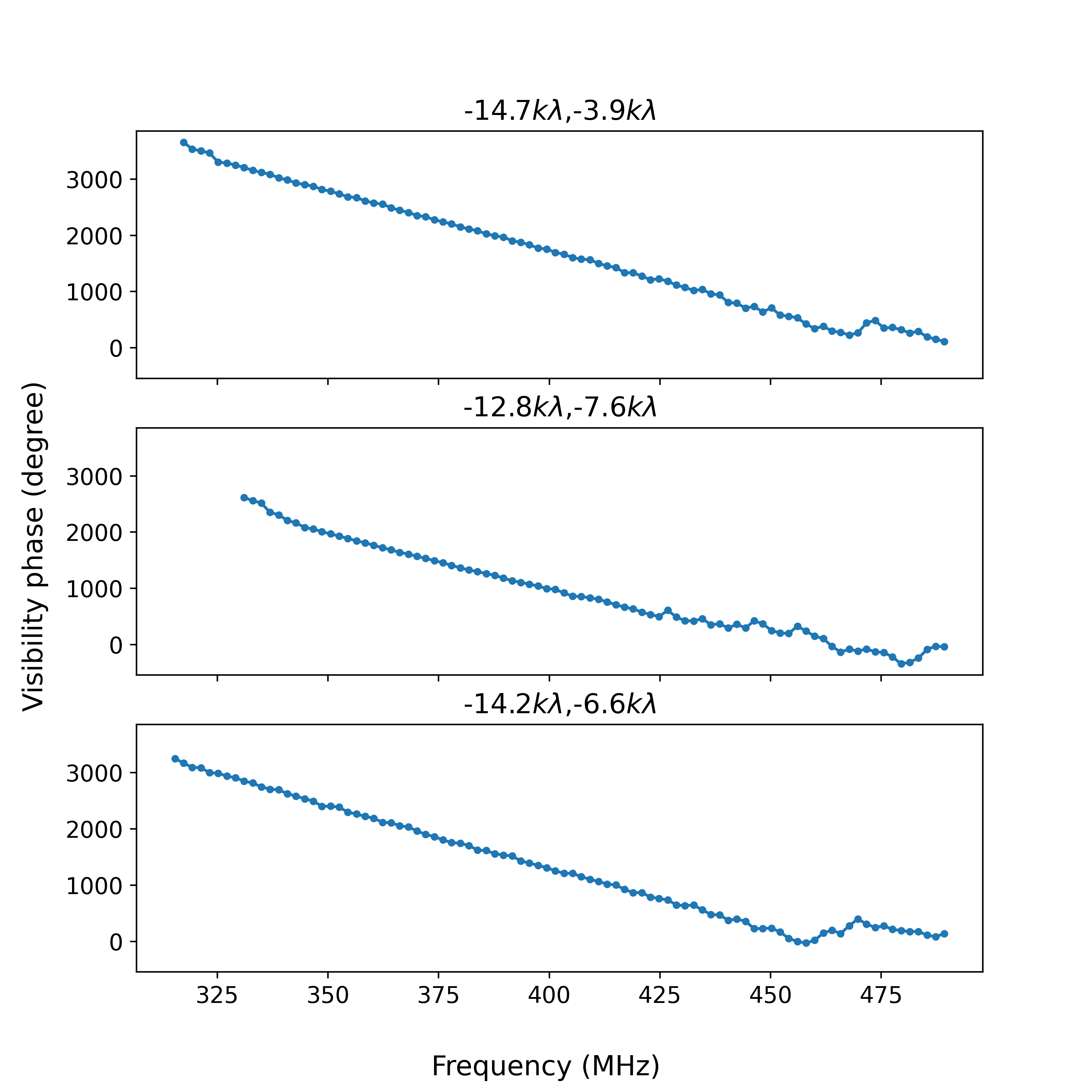}
    \caption{The variation of the visibility phase as a function of frequency is shown for three chosen baselines, using the June 4 data. The median $(u,v)$ for each baseline is mentioned on top of each panel. The median is taken over time, as each baseline samples over different $(u,v)$ points due to the rotation of the Earth.}
    \label{fig:baseline_phase_with_freq}
\end{figure}

As further evidence of the presence of these very small source sizes, in Figure \ref{fig:baseline_phase_with_freq} we show the variation of the visibility phase as a function of frequency for three chosen baselines. {The phases have been unwrapped using standard functions implemented in Numpy \citep{Harris2020}.} We have only used visibilities for which $\sqrt{u^2+v^2}/\lambda > 12000$, where $\lambda$ is the observation wavelength. The median $(u,v)$ is mentioned above each panel. The visibilities have been averaged over $\sim 2$ MHz and 25 minutes to increase the signal-to-noise ratio (SNR). As mentioned earlier, the source morphology is very similar across the 166 MHz bandwidth. Hence, we can expect that for a baseline whose length scale matches that of the source, the phase will show a linear variation with frequency for a source away from the phase centre. It is evident that the three baselines shown in Figure \ref{fig:baseline_phase_with_freq} are almost parallel to each other (evident from their median $(u,v)$, which is indicated in the title of each panel) and show a coherent linear phase variation between $\sim 325$ and $480$ MHz. This demonstrates that a source with size $\sim 13^"$ was present over the entire 25 minutes and between 325--480 MHz, same as that observed in the images.

\subsection{June 7, 2024} \label{subsec:june7}

\begin{figure}
\centering
    \includegraphics[trim={0cm 0cm 3cm 1.5cm},clip,width=\linewidth]{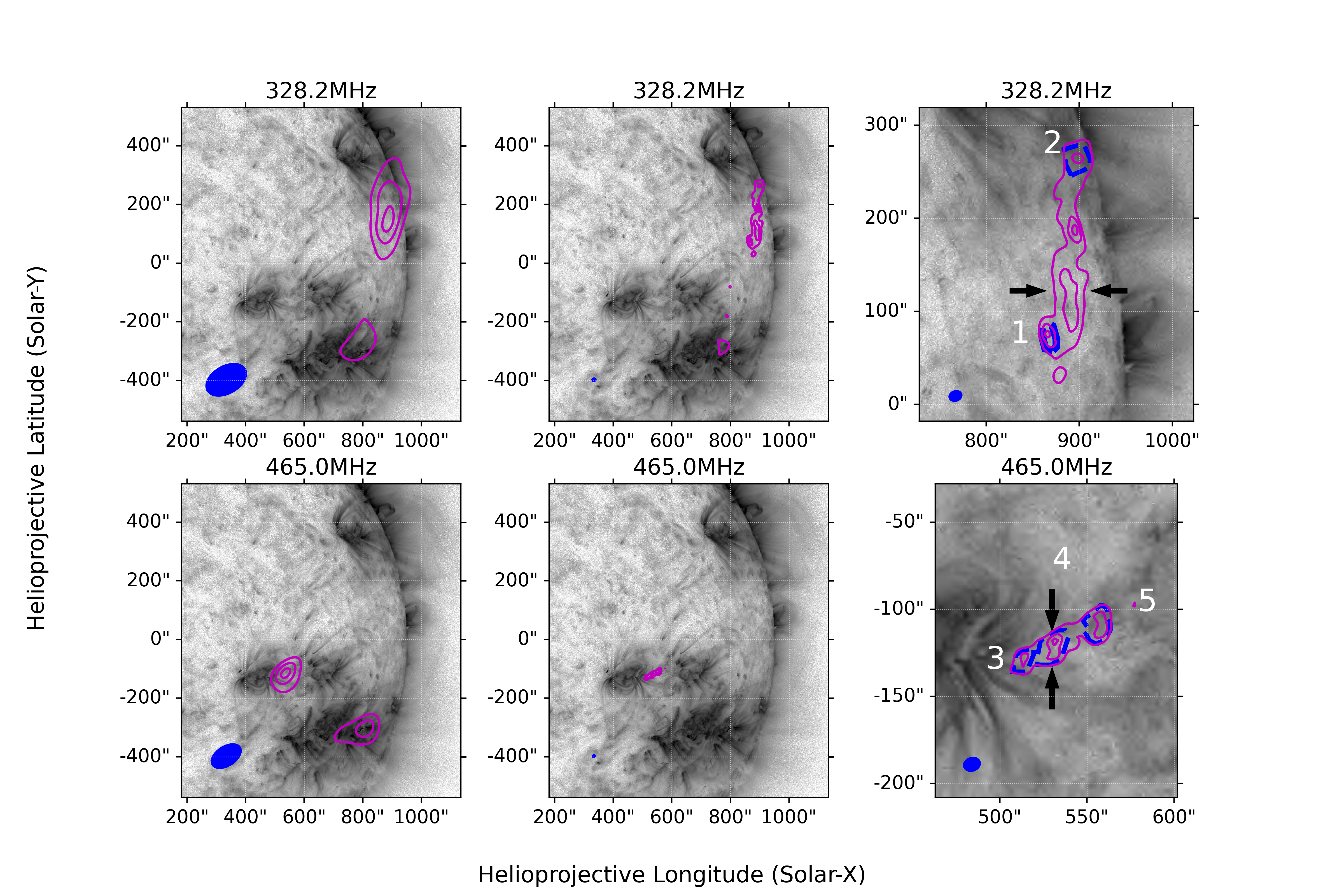}
    \caption{Images from June 7, 2024. Contours of radio images at mentioned frequencies have been overlaid on top of a AIA 131\AA$\,$ image. The contour levels are at -0.6, 0.6 0.8 and 0.95 times the peak in the respective radio images. Solid and dashed contours denote positive and negative contours, respectively.  { The positive contour values increases from the outer contours towards the inner contours.} The left column images were produced using baselines between $0.3-1k\lambda$ with \textsf{uniform} weighting. The middle panel images were produced using baselines $1-10k\lambda$ with \textsf{Briggs} weighting with \textsf{robust} value 0.5. Right panel shows zoomed in views of region of interest in the middle panel images. {All images were generated by averaging about 137 MHz and 25 minutes of data.} }
    \label{fig:june7_low_high_res}
\end{figure}

{ We observe from Figure \ref{fig:dynamic_spectrum} that the flux density of the noise storm was much smaller than that in June 4 data.} Hence, we were forced to use significant spectral averaging to make reliable images. In Figure \ref{fig:june7_low_high_res}, we show some example images. The images were generated by averaging about 137 MHz and 25 minutes of data (between 09:51--10:17 UT). The central frequency is mentioned on top of each panel. The upper and lower panels show images at two adjacent spectral bands. The left column images were produced using baselines between $0.3-1k\lambda$ with \textsf{uniform} weighting. The images in the middle panel were produced using baselines $1-10k\lambda$ with \textsf{Briggs} weighting with \textsf{robust} value 0.5. In the right panel, we have zoomed in to the location of the source of interest in the middle panel. Due to the difficulty in producing the high-resolution 328 MHz image, we started its deconvolution by providing the low-resolution image at the same frequency as the starting model. The contour levels are at -0.6, 0.6 0.8, and 0.95 times the peak in the respective radio images. Solid and dashed contours have been used to denote positive and negative contours, respectively. Adding baselines with lengths exceeding $10k\lambda$ increased the noise in the image, without any additional structure appearing, indicating that structures at angular scales smaller than $\sim 20^"$ were absent or lay below the detection thresholds.

We find that both at the high and low frequencies shown here, the radio sources are very extended along one direction,and quite compact along the orthogonal direction. The angular extents along the arrows marked in the top right and bottom right panels are approximately $30^"$ and $20^"$, respectively. In both the panels, we observe multiple compact sources inside the large extended structure. The presence of this extended structure makes it hard to model the compact sources as Gaussians. However, in the spirit of making the most of the data, we have fitted the compact sources using Gaussians using the task \textsf{imfit}. The fitted regions are shown with blue dashed lines. The convolved source sizes obtained for sources 1 and 2 in the top right panel are $18(5)^" \times 9(1)^"$ and $21(2)^" \times 19(1)^"$ respectively. The convolved source sizes obtained for the sources 3, 4 and 5, in the bottom right panel are $15(1)^"\times 8.2(0.5)^"$, $14(5)^" \times 6(1)^"$ and $14(4)\times 6.2(0.9)^"$. The size of the restoring beam is $9.6^" \times 7.6 ^"$. In all cases except source 2, \textsf{imfit} fails to find the deconvolved size as the sources are only marginally resolved along one dimension. The deconvolved source size for source 2 is $18(2)^" \times 13(2)^"$. { However, the source sizes obtained in this manner are only indicative of the true size, because of the assumption that the observed morphology of the source (which is the convolution of the restoring beam and the true source) can be well described by a Gaussian, which is not assured to be true. While the fitting algorithm does simultaneously solve for a constant offset and a Gaussian, for extended sources, it is not known a-priori, if the large scale structure, can indeed be described by a offset. The constant offset is more relevant for a noise background and is hence not necessarily true for a source with complex morphology.}

Left panel of Figure \ref{fig:june7_low_high_res} shows that among the two sources observed in the two different frequencies, only one source, located at $\sim (800^",-300^")$, is common. Both the location and the orientation of the second source are very different between the two frequencies shown. Due to this, we do not expect a linear visibility phase variation with frequency similar to that seen in Figure \ref{fig:baseline_phase_with_freq}. 

\subsection{June 11, 2024} \label{subsec:june11}

\begin{figure}
\centering
    \includegraphics[trim={0cm 0cm 3cm 1.5cm},clip,width=\textwidth]{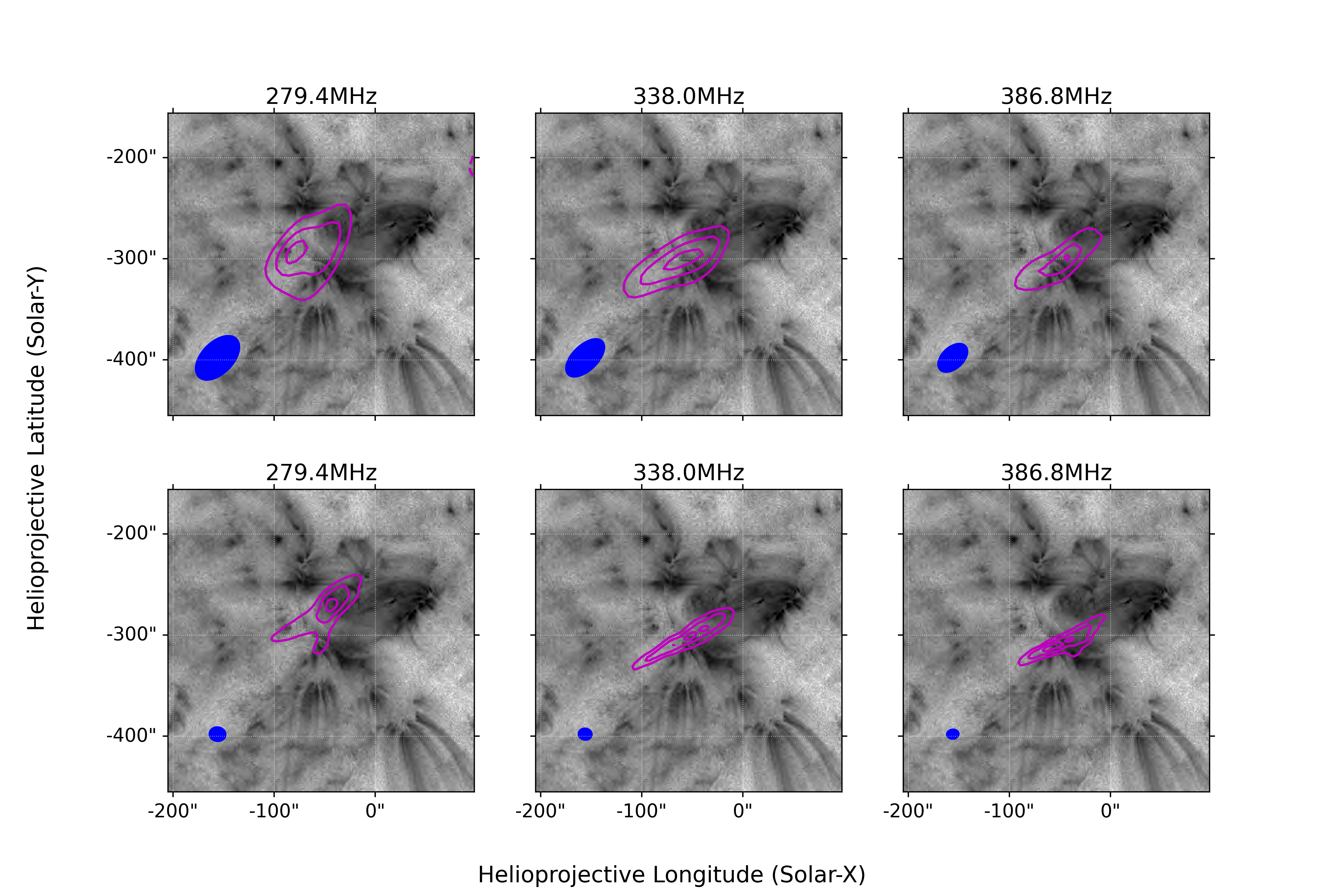}
    \caption{Image made using data from June 11, 2024. The upper and lower panels shows the low- and high-resolution radio contours overlaid on a nearby AIA 131$\AA\,$ image. The contour levels are at -0.6, 0.6 and 0.8 times the peak. Negative and positive contours are shown with dashed and solid lines, respectively.  { The positive contour values increases from the outer contours towards the inner contours.} The blue ellipse at the bottom right shows the instrumental resolution. Left and middle panel images were produced using the time interval between 09:41--10:06 UT. Images in the right panel were produced using the time interval 10:08--10:33 UT. 
    All figures have a bandwidth of 38 MHz, and the central frequencies are mentioned in title of each panel.{ The low and high resolution images are generated using baselines with lengths between $0.4-4k\lambda$ and $1-10k\lambda$ respectively. \textsf{Uniform} and \textsf{Briggs} weighting (\textsf{robust} of 0.0) were used to generate the low and high resolution images respectively.}
    }
    \label{fig:june11_low_high_res}
\end{figure}

\begin{figure}
\centering
    \includegraphics[trim={0cm 0cm 2cm 1cm},clip,width=\linewidth]{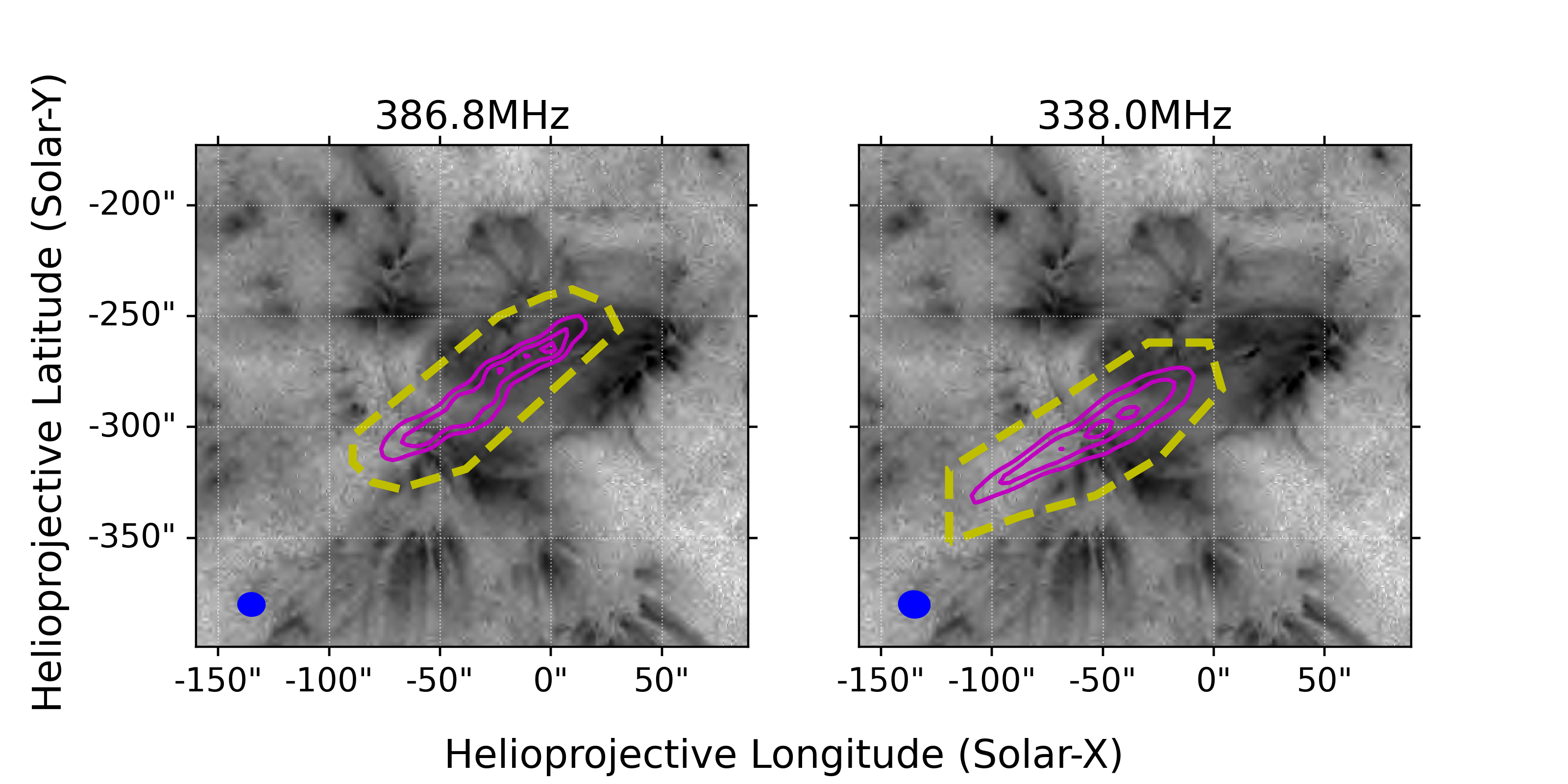}
    \caption{High resolution images are shown for two instances from June 11, 2024. The yellow dashed lines mark the region used for the Gaussian fitting.  Both images were generated using baselines between $1-10k\lambda$, with \textsf{Briggs} weighting and \textsf{robust} 0.0.}
    \label{fig:june11_highest_res}
\end{figure}

The upper and lower panels of Figure \ref{fig:june11_low_high_res} show the low- and high-resolution radio contours overlaid on AIA 131$\AA\,$ image at the closest available time. The low-resolution radio images were generated using baselines with lengths between $0.4-4k\lambda$. The high-resolution images were generated using baselines in the range $1-10k\lambda$. {The low-resolution images were generated using a \textsf{uniform} weighting scheme, whereas the high resolution images were generated using the \textsf{Briggs} weighting scheme with \textsf{robust} value of 0.0.} Approximately 19.5 MHz and 25 minutes (9:41--10:07 UT) were averaged to generate the images in the left and middle panels. The right panel images were generated using 38 MHz and 25 minutes of data (10:08--10:33 UT). The contour levels are at -0.6, 0.6 and 0.8 times the peak. Negative and positive contours are shown with dashed and solid lines, respectively. It is clear that although the detailed spatial structures of the radio source vary across this wide frequency range, qualitatively they are very similar, particularly in the low resolution images. Due to its large spectral and temporal span, we regard this as a noise storm. In Figure \ref{fig:june11_highest_res}, we show two example high-resolution images for quantitatively demonstrating the structure at small angular scales. {Both images were generated using baselines between $1-10k\lambda$, with \textsf{Briggs} weighting and \textsf{robust} 0.0.} The yellow dashed polygon shows the region which has been used to fit a Gaussian and determine the deconvolved size of the sources present inside the circle. The convolved source size of the images in the left and right panels are $126(18)^" \times 18(2)^"$ and $108(12)^" \times 22(2)^"$ respectively. The deconvolved source sizes are $120(18)^" \pm 15(3)^"$ and $102(12)^" \pm 18(2)^"$ respectively. Hence the minor axis of both sources translates to $15^" \pm 2^"$ and $18^" \pm 2^"$.

\section{Discussion} \label{sec:discussion}

\citet{kontar2019} compiled source sizes of Type III radio bursts across a wide range of frequencies between $\sim 0.04-400$ MHz and found that the observed source size in degrees can be described well by a powerlaw, given by $11.8\times f^-0.98$. On the other hand, for Type I noise storm sources, there are multiple imaging observations, where their reported full-width-half-maxima are much smaller than those expected from this powerlaw model \citep{lang1987, kerdraon1988,zlobec1992,mondal2024_noise_storm, habbal1989,mercier2015}. Barring the result of \citet{mugundhan2018}, we are unaware of any Type III source size which violates this model as significantly. However, being limited to a single baseline, \citet{mugundhan2018} were unable to image the Type IIIb source and hence their results are not as conclusive as other works. This suggests that noise storms can be more likely to show a smaller source size than Type III radio bursts. Assuming this hypothesis to be true, we can also argue that this also suggests that this difference cannot be purely because of variations in properties of the background corona, and has to be something intrinsic to the source itself. 
It is widely believed that type IIIs originate in open magnetic field lines, while type I noise storms are born in closed loops \citep[e.g.][]{mohan2021}. 
In Section \ref{subsec:scattering} we make the first attempt using a simple-minded model to understand if the closed loop can play a role in explaining the small source sizes for noise storms and also why type-IIIs generally have a larger source size.  

\subsection{Effect of Scattering in Presence of Closed Magnetic Field Loops} \label{subsec:scattering}

\begin{figure}
    \centering
    \includegraphics[trim={2cm 5cm 10cm 4cm},clip,width=0.7\linewidth]{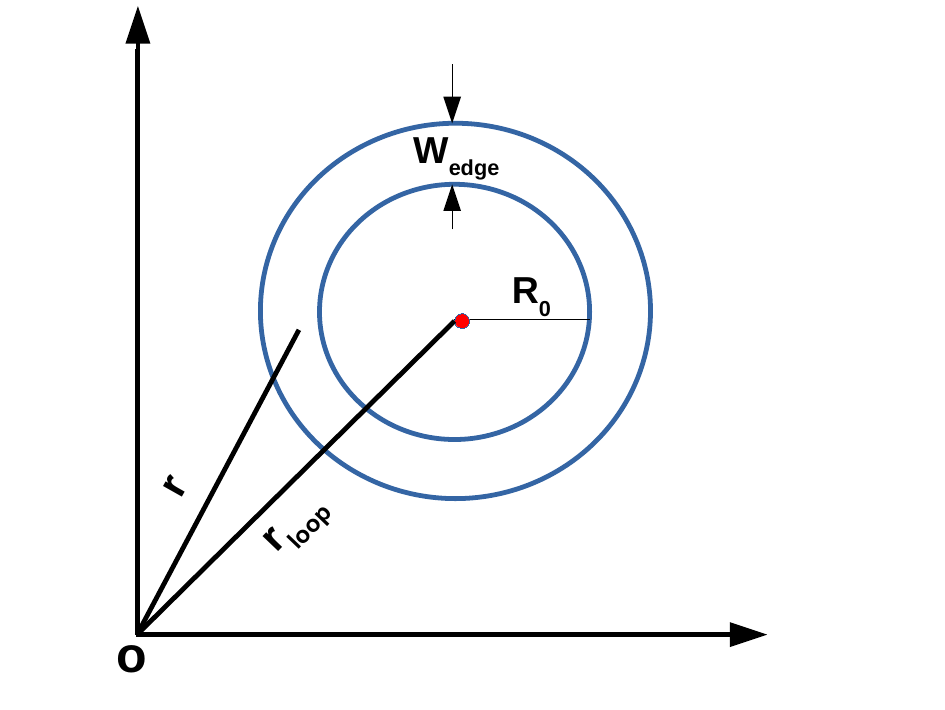}
    \caption{Shows the geometry of the coronal loop along a 2D cross-section through the centre. The loop itself is bounded by the two concentric blue circles, centered at the red circle. O denotes the origin of the coordinate system.}
    \label{fig:coronal_loop}
\end{figure}
\begin{figure}
    \centering
    \includegraphics[width=0.435\linewidth]{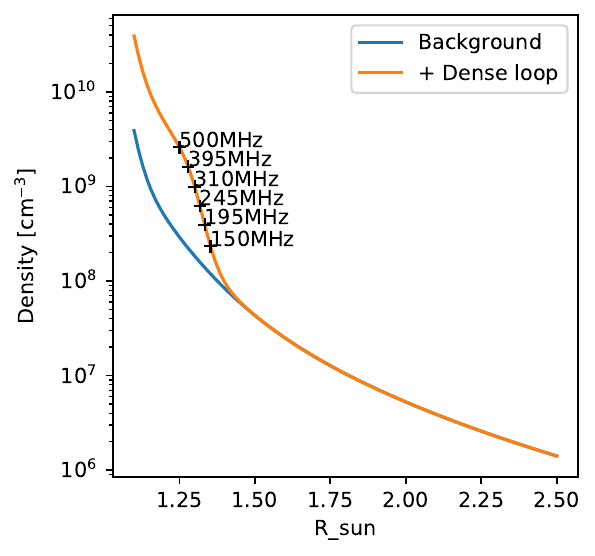}
    \includegraphics[width=0.52 \linewidth]{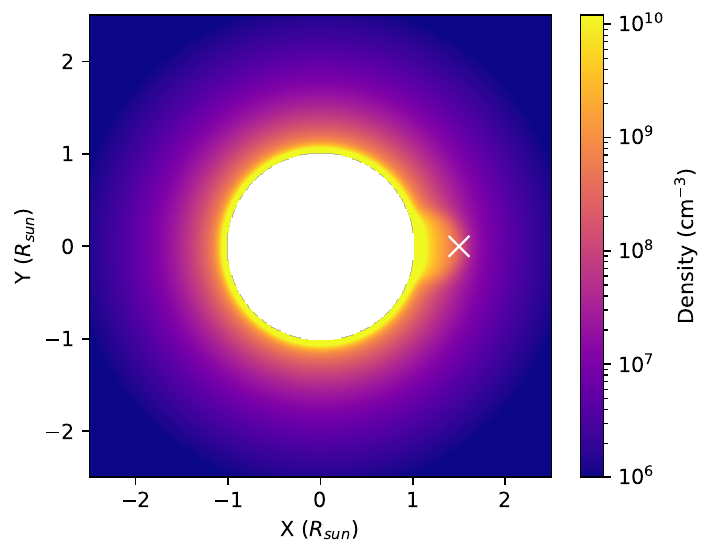}
    \caption{Left panel shows the variation of density with heliocentric distance. The cross-mark beside each frequency indicates the emission heights, assuming the emission is happening at the second harmonic of the plasma frequency.  Right panel shows the plasma density variation when the over-dense region is placed at the western limb of the Sun. }
    \label{fig:densDist}
\end{figure}
\begin{figure}
    \centering
    \includegraphics[width=0.5\linewidth]{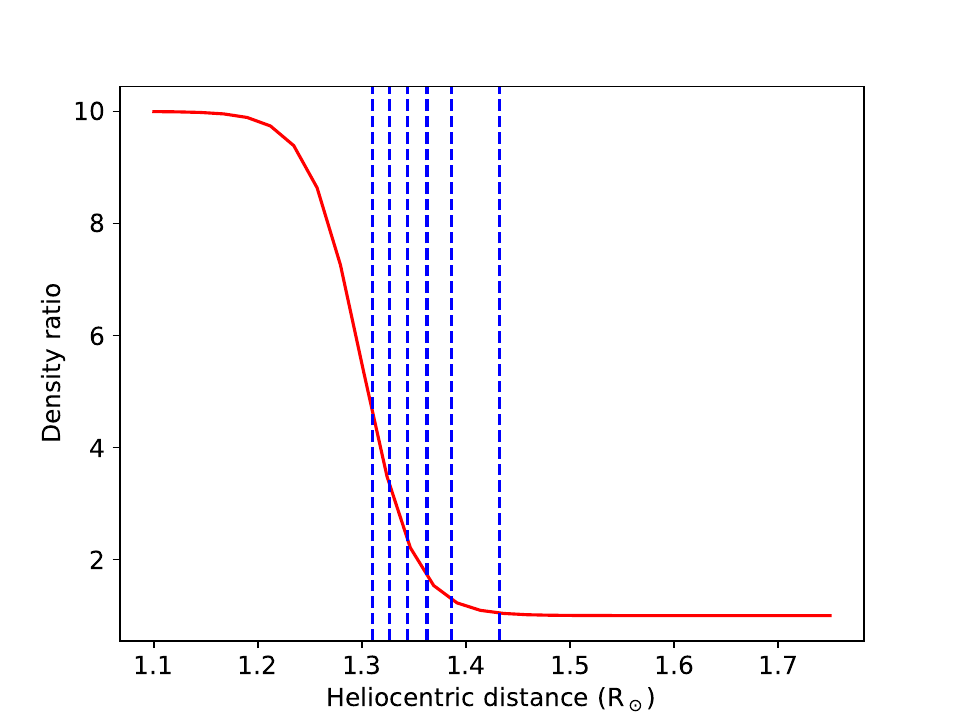}
    \caption{Shows the ratio between the background density and the density in the coronal loop. The dashed lines from left to right shows the locations where the emission at frequencies of 500, 395, 310, 245, 195, 150 MHz originates from (assuming harmonic emission). }
    \label{fig:density_ratio}
\end{figure}

In this work, we performed radio wave propagation simulations modeling a radio source originating in an over-dense loop top to see how the large-scale density structure influences the observed source size. The simulation uses the ray tracing and scattering modeling method from \citet{zhang2021parametric}, which implements the model developed in \citet{kontar2019} and  \citet{kontar2023anisotropic}. The background density is modified to include the density of the over-dense loop, $R_+$,
\begin{equation}
   N_e (\vec{r})
= N_{e0}(\vec{r}) \times (1+R_+).
\end{equation}
where:
\begin{equation}
   R_+= \bigl(R_N - 1\bigr)\,
\frac{1 - \tanh\bigl(( |\vec{r}- \vec{r}_{loop}| - R_{0})/\,W_{edge}\bigr)}{2}
\end{equation}
$N_{e0}(\vec{r})$ is the background density distribution in a spherically symmetric classical density model, same as used by \citet{kontar2019}.
$R_N$ is the ratio of the density compared to $N_{e0}$, $\vec{r}_{loop}$ is the vector location of the center of the over dense region, $R_0$ is the radius of the region (we assume that the over-dense region has a spherical geometry), $W_{edge}$ is the width of the over-dense region. We have chosen this particular form due to the asymptotic property of the $\tanh$, which ensures that when $|\vec{r}|>>|\vec{r_{loop}}|$, there is no contribution from the loop model. A cartoon diagram of the coronal loop model, showing the geometric parameters is shown in Figure \ref{fig:coronal_loop}.
{ We used parameter set of $R_N = 10$, $R_0=1.1$, $r_{\rm loop}=0.2$, and $W_{\rm edge}=0.05$, as demonstration and implemented it in the following wave propagation modeling}.
The composite density distribution is presented in Figure \ref{fig:densDist}, where we have placed the dense coronal loop at the western limb. We have performed ray-tracing simulations, with and without the over-dense loop, for six frequencies -- 150, 195, 245, 310, 395, and 500 MHz. { This is motivated by the fact that the effect of scattering and consequently the scattered source size is expected to increase with $\omega_{pe}/\omega$, where $\omega_{pe},\ \omega$ are the local plasma frequency and observation frequency respectively. } In Figure \ref{fig:density_ratio} we show the ratio between the density inside the coronal loop and the background density. The dashed lines represent the emission heights corresponding to 150, 195, 245, 310, 395, and 500 MHz, which are the same ones indicated in Figure \ref{fig:densDist}, with the highest frequency increasing at lowest height. The emission height increases with decrease in frequency. In these simulations, we have assumed that the emission frequency is the harmonic of the local plasma frequency. {If the observed emission is at the fundamental mode, explaining the observed source sizes becomes even more challenging. }


\begin{figure}
    \centering
    \includegraphics[width=0.33\linewidth]{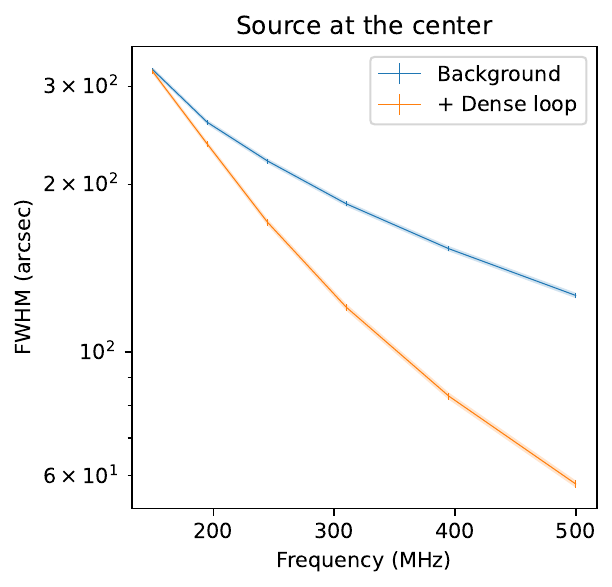}
    \includegraphics[width=0.31\linewidth]{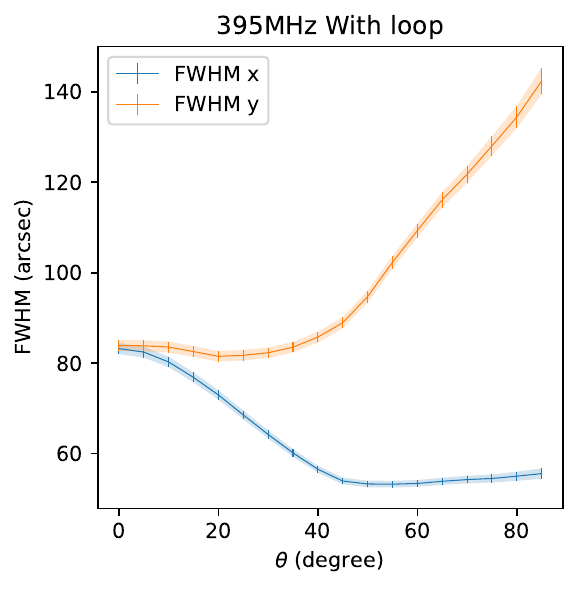}
    \includegraphics[width=0.31\linewidth]{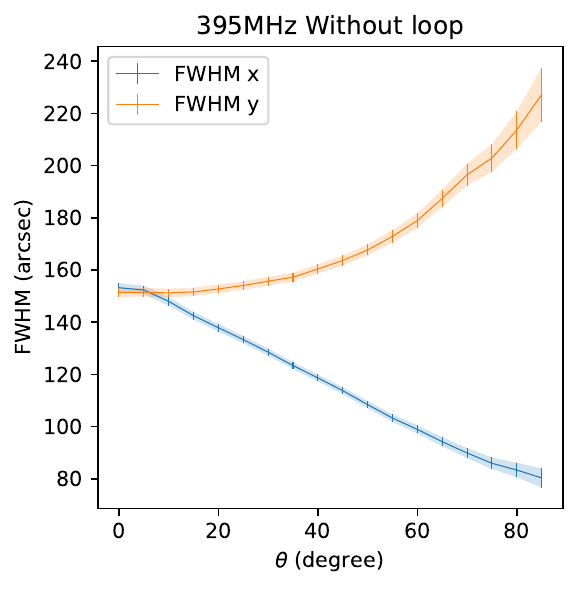}
    \caption{Left panel shows the variation of source size with the observing frequency. The orange line is used to show the source size variation when the over-dense region is present at the disc center. Blue line is used to denote the case when no over-dense region is present. Middle and right panels show the source size variation with the heliocentric longitude of the emission source at 395 MHz. The FWHMs at mutually orthogonal directions have been shown with different colors in each panel. X and Y refers to the horizontal and vertical directions and are shown in right panel of Figure \ref{fig:densDist}.}
    \label{fig:angFWHM}
\end{figure}

In the left panel of Figure \ref{fig:angFWHM}, we show the variation of source size with the observing frequency for two cases: a) without an over-dense loop, b) with an over-dense loop at the solar center. The simulation tool used here provides an image of the source, which was then fitted with a gaussian to obtain the full-width half-maximums (FWHMs) along the two orthogonal directions. It is evident that at frequencies higher than 200 MHz, the size of the source is significantly reduced for the source originating from the over-dense region. For frequencies higher than 400 MHz, the source size from the non-over-dense case is twice that for  the over-dense case. However, the presence of the over-dense loop has  little effect at frequencies below 200 MHz.  In the middle and right panels of Figure \ref{fig:angFWHM}, we have explored the effect of the location of the over-dense with respect to the disc center, where both axes of the coronal loop are perpendicular to the line-of-sight. We find that the minor axis FWHM of the source has a trend decreasing toward the limb of the Sun. This trend is seen for simulations both with and without over-dense region, and serves as a consistency check for these simulations \citep{kontar2019}. To summarize, these simulations indicate that the existence of an over-dense region can impact the source size significantly, and under favorable circumstances can reduce the angular scales associated with a source by as much as $\sim$3.5 times the generic case with no over-dense region. 

\subsection{Caveats and Future Direction for Modeling}

{The minimum source angular size obtained from these simulations is approximately $50^{"}$. In comparison, the minimum source observed source size at a comparable frequency (Figure \ref{fig:june11_highest_res}) is $\sim 20^{"}$, which is more than half the minimum predicted size. While this discrepancy between the predicted and observed source size points to model limitations, the simulations presented in this work provide us with some valuable insights.} 
Through this work, we have demonstrated that if the emission source is located within an over-dense region, the observed source size can be much smaller than that expected for the generic case, i.e. in absence of any density enhancements. This  can be understood based on the physical picture for scattering. Given the rapid drop in coronal electron density with radius, most of the scattering happens close to the region where the plasma frequency is comparable to the emission frequency. When the emission source is embedded inside an over-dense region, while still inside, it encounters a comparable plasma frequency, corresponding to the over-dense region. Once the light ray exits the over-dense structure, it abruptly finds itself in a region with much lower plasma frequency and hence suffers much less additional scattering for the rest of its path. 
The path length in the over-dense region is effectively determined by the size of the higher-density structure, which would usually be much smaller than the density scale height of the background coronal density. Hence, the effective path length over which significant scattering occurs in the generic case, is much larger than when the emission arises in an over-dense region. This also explains why we did not see much effect of the composite model at 150 MHz in the left panel of Figure \ref{fig:angFWHM}. The density distribution seen in the left panel of Figure \ref{fig:densDist} shows that the density contrast between the composite model and the background model is much small at the emission height of 150 MHz, than at 310 MHz. Although, this work could not reproduce the observed source size, it does make the discrepancy much smaller. Thus this limited initial exploration already points to  a potential solution towards understanding the origin of such small source sizes observed in case of noise-storms, which are known to originate from coronal loops.

However, there are several opportunities for improvement in this modeling approach. First, the simulation software assumes that the magnetic field lines are purely radial. While this may hold true at higher coronal heights, for which this simulation tool was primarily used, this is not valid in the low corona, where such over-dense structures with non-radial magnetic field can be present. Theoretical and simulation works suggest that the density inhomogeneities are elongated along the magnetic field \citep{kontar2019,kontar2023anisotropic}, and hence the scatter-broadening is less if the LOS is aligned with the magnetic field direction. Hence, it is natural to expect that the orientation of the magnetic field inside the coronal loop will affect the scattering, depending on the angle between the LOS and the local magnetic field direction. For example, FWHM along the X-axis is much smaller when the over-dense region is located at the limb, because at this orientation, the magnetic field direction and the X-axis are aligned with each other. Such behavior can also be seen depending on the orientation of the coronal loop, which can decrease the observed length scales even further, than that achieved using this illustrative model. To the best of our knowledge, \citet{mohan2021} provides the only available estimate estimated of the heliocentric dependence in the low corona, of the relative density fluctuation variance ($\epsilon$), given by $<\delta n^2>/n^2$, where $n$ and $\delta n$ are the background density and density inhomogeneity respectively. However, simulating the scatter-broadened source sizes requires knowledge of $\bar{q}\epsilon^2$, where $\bar{q}$ is the normalized first moment of the turbulence spectrum in the solar corona \citep{kontar2019}, which is not available at these heights. We have used measurements of the anisotropy level and density inhomogeneities derived from much higher coronal heights \citep{kontar2023anisotropic}.  These values might not hold at these low coronal heights, potentially affecting the results presented here. 

\section{Conclusion} \label{sec:conclusion}
Here we have presented multiple instances where the observed type-I noise storms with source sizes smaller than $20^"$. 
For the data from June 4, 2024, we have provided additional evidence for the presence of structures with length scales $\lesssim 13^"$. This doubles the number of instances where such small source sizes have been detected. M24 pointed out that all the previous detections were noise storms located close to the limb. This work demonstrates that to merely be sample bias. Both on June 4 and June 11, the detected noise storm was located close to the disc center, and still we found the presence of sources with small angular-scales in them. Scattering theories have generally predicted the impossibility of observing such small scale sources, based on  the level of small scale inhomogeneities observed in the solar corona. However, the mounting evidence demonstrating the presence of small-angular scale sources suggests that the conditions under which such compact structures can be observed in type-I noise storms cannot be rare. Additionally, since in all of these cases, we have averaged over approximately 15--20 minutes, it rules out the possibility that the source sizes are a result of chance alignments of the density inhomogeneities in the solar corona. 

We have also taken the first steps towards building an understanding of the presence of these small angular scales observed in noise storms. We have demonstrated, using a simple-minded model, that the presence of over-dense structures in the emitting region, can significantly reduce the observed source size, and might be responsible for the small sources observed here and in M24. Although our simulations fall short of reproducing the observed source sizes, they substantially reduce the differences between predicted and observed source sizes.
This strong suggests that this a good direction to pursue for understanding the existence of these small coronal radio sources.

\section{Funding}

The Department of Atomic Energy, Government of India, under project no. 12-R\&D-TFR-5.02-0700 provided partial funding to DO for this work. P. Z. was supported by the NASA Living with a Star Jack Eddy Postdoctoral Fellowship Program, administered by UCAR’s Cooperative Programs for the Advancement of Earth System Science (CPAESS) under award 80NSSC22M0097 for this research.

\begin{acks}
    This work uses observations from the Giant Metrewave Radio Telescope (GMRT) run by the National Centre for Radio Astrophysics of the Tata Institute of Fundamental Research. We thank the staff of the GMRT, 
especially the observing staff, 
who made these non-trivial solar observations possible. 
D.O. acknowledges the support of the Department of Atomic Energy, Government of India, under project no. 12-R\&D-TFR-5.02-0700. 
P. Z. acknowledges support for this research by the NASA Living with a Star Jack Eddy Postdoctoral Fellowship Program, administered by UCAR’s Cooperative Programs for the Advancement of Earth System Science (CPAESS) under award 80NSSC22M0097.
The authors gratefully acknowledges the developers of CASA \citep{CASA2022}, SunPy \citep{sunpy_community2020,sunpy2.1.0}, astropy \citep{astropy:2013,astropy:2018,astropy:2022}, matplotlib \citep{Hunter:2007}, Numpy \citep{Harris2020}, Scipy \citep[][]{Scipy2020}, Python 3 \citep[][]{python3}.

\end{acks}

\bibliographystyle{spr-mp-sola}
\bibliography{sola_bibliography_example}{}

\begin{thebibliography}{44}
\ifx\bisbn     \undefined \def\bisbn  #1{ISBN #1}\fi
\ifx\binits    \undefined \def\binits#1{#1}\fi
\ifx\bauthor   \undefined \def\bauthor#1{#1}\fi
\ifx\batitle   \undefined \def\batitle#1{#1}\fi
\ifx\bjtitle   \undefined \def\bjtitle#1{\textit{#1}}\fi
\ifx\bvolume   \undefined \def\bvolume#1{\textbf{#1}}\fi
\ifx\byear     \undefined \def\byear#1{#1}\fi
\ifx\bissue    \undefined \def\bissue#1{#1}\fi
\ifx\bfpage    \undefined \def\bfpage#1{#1}\fi
\ifx\blpage    \undefined \def\blpage #1{#1}\fi
\ifx\burl      \undefined \def\burl#1{\textsf{#1}}\fi
\ifx\href      \undefined \def\href#1#2{\textsf{#2}}\fi
\ifx\betal     \undefined \def\betal{\textit{et al.}}\fi
\ifx\bctitle   \undefined \def\bctitle#1{#1}\fi
\ifx\beditor   \undefined \def\beditor#1{#1}\fi
\ifx\bbtitle   \undefined \def\bbtitle#1{\textit{#1}}\fi
\ifx\bedition  \undefined \def\bedition#1{#1}\fi
\ifx\bseriesno \undefined \def\bseriesno#1{\textbf{#1}}\fi
\ifx\blocation \undefined \def\blocation#1{#1}\fi
\ifx\bsertitle \undefined \def\bsertitle#1{\textit{#1}}\fi
\ifx\bsnm      \undefined \def\bsnm#1{#1}\fi
\ifx\bsuffix   \undefined \def\bsuffix#1{#1}\fi
\ifx\bparticle \undefined \def\bparticle#1{#1}\fi
\ifx\barticle  \undefined \def\barticle#1{}\fi
\ifx\binstitute  \undefined \def\binstitute#1{#1}\fi
\ifx\bpublisher  \undefined \def\bpublisher#1{#1}\fi
\ifx\doiurl    \undefined
  \def\doiurl#1{\href{http://dx.doi.org/#1}{\textsf{DOI}}}\fi
\ifx\arxivurl  \undefined
  \def\arxivurl#1{\href{http://arxiv.org/abs/#1}{\textsf{arXiv}}}\fi
\ifx\adsurl    \undefined
  \def\adsurl#1{\href{http://adsabs.harvard.edu/abs/#1}{\textsf{ADS}}}\fi
\ifx\botherref \undefined \def\botherref#1{}\fi
\ifx\url       \undefined \def\url#1{\textsf{#1}}\fi
\ifx\bchapter  \undefined \def\bchapter#1{}\fi
\ifx\bbook     \undefined \def\bbook#1{}\fi
\ifx\bcomment  \undefined \def\bcomment#1{#1}\fi
\ifx\oauthor   \undefined \def\oauthor#1{#1}\fi
\ifx\citeauthoryear \undefined\def \citeauthoryear#1{#1}\fi
\ifx\endbibitem\undefined \def\endbibitem{}\fi
\ifx\bconflocation  \undefined \def\bconflocation#1{#1} \fi

\bibitem[\protect\citeauthoryear{{Astropy Collaboration}
  {et~al.}}{2013}]{astropy:2013}
\begin{barticle}
\bauthor{\bsnm{{Astropy Collaboration}}},
\bauthor{\bsnm{{Robitaille}}, \binits{T.P.}},
\bauthor{\bsnm{{Tollerud}}, \binits{E.J.}},
\bauthor{\bsnm{{Greenfield}}, \binits{P.}},
\bauthor{\bsnm{{Droettboom}}, \binits{M.}},
\bauthor{\bsnm{{Bray}}, \binits{E.}},
\bauthor{\bsnm{{Aldcroft}}, \binits{T.}},
\bauthor{\bsnm{{Davis}}, \binits{M.}},
\bauthor{\bsnm{{Ginsburg}}, \binits{A.}},
\bauthor{\bsnm{{Price-Whelan}}, \binits{A.M.}},
\bauthor{\bsnm{{Kerzendorf}}, \binits{W.E.}},
\bauthor{\bsnm{{Conley}}, \binits{A.}},
\bauthor{\bsnm{{Crighton}}, \binits{N.}},
\bauthor{\bsnm{{Barbary}}, \binits{K.}},
\bauthor{\bsnm{{Muna}}, \binits{D.}},
\bauthor{\bsnm{{Ferguson}}, \binits{H.}},
\bauthor{\bsnm{{Grollier}}, \binits{F.}},
\bauthor{\bsnm{{Parikh}}, \binits{M.M.}},
\bauthor{\bsnm{{Nair}}, \binits{P.H.}},
\bauthor{\bsnm{{Unther}}, \binits{H.M.}},
\bauthor{\bsnm{{Deil}}, \binits{C.}},
\bauthor{\bsnm{{Woillez}}, \binits{J.}},
\bauthor{\bsnm{{Conseil}}, \binits{S.}},
\bauthor{\bsnm{{Kramer}}, \binits{R.}},
\bauthor{\bsnm{{Turner}}, \binits{J.E.H.}},
\bauthor{\bsnm{{Singer}}, \binits{L.}},
\bauthor{\bsnm{{Fox}}, \binits{R.}},
\bauthor{\bsnm{{Weaver}}, \binits{B.A.}},
\bauthor{\bsnm{{Zabalza}}, \binits{V.}},
\bauthor{\bsnm{{Edwards}}, \binits{Z.I.}},
\bauthor{\bsnm{{Azalee Bostroem}}, \binits{K.}},
\bauthor{\bsnm{{Burke}}, \binits{D.J.}},
\bauthor{\bsnm{{Casey}}, \binits{A.R.}},
\bauthor{\bsnm{{Crawford}}, \binits{S.M.}},
\bauthor{\bsnm{{Dencheva}}, \binits{N.}},
\bauthor{\bsnm{{Ely}}, \binits{J.}},
\bauthor{\bsnm{{Jenness}}, \binits{T.}},
\bauthor{\bsnm{{Labrie}}, \binits{K.}},
\bauthor{\bsnm{{Lim}}, \binits{P.L.}},
\bauthor{\bsnm{{Pierfederici}}, \binits{F.}},
\bauthor{\bsnm{{Pontzen}}, \binits{A.}},
\bauthor{\bsnm{{Ptak}}, \binits{A.}},
\bauthor{\bsnm{{Refsdal}}, \binits{B.}},
\bauthor{\bsnm{{Servillat}}, \binits{M.}},
\bauthor{\bsnm{{Streicher}}, \binits{O.}}:
\byear{2013},
\batitle{{Astropy: A community Python package for astronomy}}.
\bjtitle{\aap}
\bvolume{558},
\bfpage{A33}.
\doiurl{10.1051/0004-6361/201322068}.
\adsurl{2013A\%26A...558A..33A}.
\end{barticle}
\endbibitem

\bibitem[\protect\citeauthoryear{{Astropy Collaboration}
  {et~al.}}{2018}]{astropy:2018}
\begin{barticle}
\bauthor{\bsnm{{Astropy Collaboration}}},
\bauthor{\bsnm{{Price-Whelan}}, \binits{A.M.}},
\bauthor{\bsnm{{Sip{\H{o}}cz}}, \binits{B.M.}},
\bauthor{\bsnm{{G{\"u}nther}}, \binits{H.M.}},
\bauthor{\bsnm{{Lim}}, \binits{P.L.}},
\bauthor{\bsnm{{Crawford}}, \binits{S.M.}},
\bauthor{\bsnm{{Conseil}}, \binits{S.}},
\bauthor{\bsnm{{Shupe}}, \binits{D.L.}},
\bauthor{\bsnm{{Craig}}, \binits{M.W.}},
\bauthor{\bsnm{{Dencheva}}, \binits{N.}},
\bauthor{\bsnm{{Ginsburg}}, \binits{A.}},
\bauthor{\bsnm{{Vand erPlas}}, \binits{J.T.}},
\bauthor{\bsnm{{Bradley}}, \binits{L.D.}},
\bauthor{\bsnm{{P{\'e}rez-Su{\'a}rez}}, \binits{D.}},
\bauthor{\bsnm{{de Val-Borro}}, \binits{M.}},
\bauthor{\bsnm{{Aldcroft}}, \binits{T.L.}},
\bauthor{\bsnm{{Cruz}}, \binits{K.L.}},
\bauthor{\bsnm{{Robitaille}}, \binits{T.P.}},
\bauthor{\bsnm{{Tollerud}}, \binits{E.J.}},
\bauthor{\bsnm{{Ardelean}}, \binits{C.}},
\bauthor{\bsnm{{Babej}}, \binits{T.}},
\bauthor{\bsnm{{Bach}}, \binits{Y.P.}},
\bauthor{\bsnm{{Bachetti}}, \binits{M.}},
\bauthor{\bsnm{{Bakanov}}, \binits{A.V.}},
\bauthor{\bsnm{{Bamford}}, \binits{S.P.}},
\bauthor{\bsnm{{Barentsen}}, \binits{G.}},
\bauthor{\bsnm{{Barmby}}, \binits{P.}},
\bauthor{\bsnm{{Baumbach}}, \binits{A.}},
\bauthor{\bsnm{{Berry}}, \binits{K.L.}},
\bauthor{\bsnm{{Biscani}}, \binits{F.}},
\bauthor{\bsnm{{Boquien}}, \binits{M.}},
\bauthor{\bsnm{{Bostroem}}, \binits{K.A.}},
\bauthor{\bsnm{{Bouma}}, \binits{L.G.}},
\bauthor{\bsnm{{Brammer}}, \binits{G.B.}},
\bauthor{\bsnm{{Bray}}, \binits{E.M.}},
\bauthor{\bsnm{{Breytenbach}}, \binits{H.}},
\bauthor{\bsnm{{Buddelmeijer}}, \binits{H.}},
\bauthor{\bsnm{{Burke}}, \binits{D.J.}},
\bauthor{\bsnm{{Calderone}}, \binits{G.}},
\bauthor{\bsnm{{Cano Rodr{\'\i}guez}}, \binits{J.L.}},
\bauthor{\bsnm{{Cara}}, \binits{M.}},
\bauthor{\bsnm{{Cardoso}}, \binits{J.V.M.}},
\bauthor{\bsnm{{Cheedella}}, \binits{S.}},
\bauthor{\bsnm{{Copin}}, \binits{Y.}},
\bauthor{\bsnm{{Corrales}}, \binits{L.}},
\bauthor{\bsnm{{Crichton}}, \binits{D.}},
\bauthor{\bsnm{{D'Avella}}, \binits{D.}},
\bauthor{\bsnm{{Deil}}, \binits{C.}},
\bauthor{\bsnm{{Depagne}}, \binits{{\'E}.}},
\bauthor{\bsnm{{Dietrich}}, \binits{J.P.}},
\bauthor{\bsnm{{Donath}}, \binits{A.}},
\bauthor{\bsnm{{Droettboom}}, \binits{M.}},
\bauthor{\bsnm{{Earl}}, \binits{N.}},
\bauthor{\bsnm{{Erben}}, \binits{T.}},
\bauthor{\bsnm{{Fabbro}}, \binits{S.}},
\bauthor{\bsnm{{Ferreira}}, \binits{L.A.}},
\bauthor{\bsnm{{Finethy}}, \binits{T.}},
\bauthor{\bsnm{{Fox}}, \binits{R.T.}},
\bauthor{\bsnm{{Garrison}}, \binits{L.H.}},
\bauthor{\bsnm{{Gibbons}}, \binits{S.L.J.}},
\bauthor{\bsnm{{Goldstein}}, \binits{D.A.}},
\bauthor{\bsnm{{Gommers}}, \binits{R.}},
\bauthor{\bsnm{{Greco}}, \binits{J.P.}},
\bauthor{\bsnm{{Greenfield}}, \binits{P.}},
\bauthor{\bsnm{{Groener}}, \binits{A.M.}},
\bauthor{\bsnm{{Grollier}}, \binits{F.}},
\bauthor{\bsnm{{Hagen}}, \binits{A.}},
\bauthor{\bsnm{{Hirst}}, \binits{P.}},
\bauthor{\bsnm{{Homeier}}, \binits{D.}},
\bauthor{\bsnm{{Horton}}, \binits{A.J.}},
\bauthor{\bsnm{{Hosseinzadeh}}, \binits{G.}},
\bauthor{\bsnm{{Hu}}, \binits{L.}},
\bauthor{\bsnm{{Hunkeler}}, \binits{J.S.}},
\bauthor{\bsnm{{Ivezi{\'c}}}, \binits{{\v{Z}}.}},
\bauthor{\bsnm{{Jain}}, \binits{A.}},
\bauthor{\bsnm{{Jenness}}, \binits{T.}},
\bauthor{\bsnm{{Kanarek}}, \binits{G.}},
\bauthor{\bsnm{{Kendrew}}, \binits{S.}},
\bauthor{\bsnm{{Kern}}, \binits{N.S.}},
\bauthor{\bsnm{{Kerzendorf}}, \binits{W.E.}},
\bauthor{\bsnm{{Khvalko}}, \binits{A.}},
\bauthor{\bsnm{{King}}, \binits{J.}},
\bauthor{\bsnm{{Kirkby}}, \binits{D.}},
\bauthor{\bsnm{{Kulkarni}}, \binits{A.M.}},
\bauthor{\bsnm{{Kumar}}, \binits{A.}},
\bauthor{\bsnm{{Lee}}, \binits{A.}},
\bauthor{\bsnm{{Lenz}}, \binits{D.}},
\bauthor{\bsnm{{Littlefair}}, \binits{S.P.}},
\bauthor{\bsnm{{Ma}}, \binits{Z.}},
\bauthor{\bsnm{{Macleod}}, \binits{D.M.}},
\bauthor{\bsnm{{Mastropietro}}, \binits{M.}},
\bauthor{\bsnm{{McCully}}, \binits{C.}},
\bauthor{\bsnm{{Montagnac}}, \binits{S.}},
\bauthor{\bsnm{{Morris}}, \binits{B.M.}},
\bauthor{\bsnm{{Mueller}}, \binits{M.}},
\bauthor{\bsnm{{Mumford}}, \binits{S.J.}},
\bauthor{\bsnm{{Muna}}, \binits{D.}},
\bauthor{\bsnm{{Murphy}}, \binits{N.A.}},
\bauthor{\bsnm{{Nelson}}, \binits{S.}},
\bauthor{\bsnm{{Nguyen}}, \binits{G.H.}},
\bauthor{\bsnm{{Ninan}}, \binits{J.P.}},
\bauthor{\bsnm{{N{\"o}the}}, \binits{M.}},
\bauthor{\bsnm{{Ogaz}}, \binits{S.}},
\bauthor{\bsnm{{Oh}}, \binits{S.}},
\bauthor{\bsnm{{Parejko}}, \binits{J.K.}},
\bauthor{\bsnm{{Parley}}, \binits{N.}},
\bauthor{\bsnm{{Pascual}}, \binits{S.}},
\bauthor{\bsnm{{Patil}}, \binits{R.}},
\bauthor{\bsnm{{Patil}}, \binits{A.A.}},
\bauthor{\bsnm{{Plunkett}}, \binits{A.L.}},
\bauthor{\bsnm{{Prochaska}}, \binits{J.X.}},
\bauthor{\bsnm{{Rastogi}}, \binits{T.}},
\bauthor{\bsnm{{Reddy Janga}}, \binits{V.}},
\bauthor{\bsnm{{Sabater}}, \binits{J.}},
\bauthor{\bsnm{{Sakurikar}}, \binits{P.}},
\bauthor{\bsnm{{Seifert}}, \binits{M.}},
\bauthor{\bsnm{{Sherbert}}, \binits{L.E.}},
\bauthor{\bsnm{{Sherwood-Taylor}}, \binits{H.}},
\bauthor{\bsnm{{Shih}}, \binits{A.Y.}},
\bauthor{\bsnm{{Sick}}, \binits{J.}},
\bauthor{\bsnm{{Silbiger}}, \binits{M.T.}},
\bauthor{\bsnm{{Singanamalla}}, \binits{S.}},
\bauthor{\bsnm{{Singer}}, \binits{L.P.}},
\bauthor{\bsnm{{Sladen}}, \binits{P.H.}},
\bauthor{\bsnm{{Sooley}}, \binits{K.A.}},
\bauthor{\bsnm{{Sornarajah}}, \binits{S.}},
\bauthor{\bsnm{{Streicher}}, \binits{O.}},
\bauthor{\bsnm{{Teuben}}, \binits{P.}},
\bauthor{\bsnm{{Thomas}}, \binits{S.W.}},
\bauthor{\bsnm{{Tremblay}}, \binits{G.R.}},
\bauthor{\bsnm{{Turner}}, \binits{J.E.H.}},
\bauthor{\bsnm{{Terr{\'o}n}}, \binits{V.}},
\bauthor{\bsnm{{van Kerkwijk}}, \binits{M.H.}},
\bauthor{\bsnm{{de la Vega}}, \binits{A.}},
\bauthor{\bsnm{{Watkins}}, \binits{L.L.}},
\bauthor{\bsnm{{Weaver}}, \binits{B.A.}},
\bauthor{\bsnm{{Whitmore}}, \binits{J.B.}},
\bauthor{\bsnm{{Woillez}}, \binits{J.}},
\bauthor{\bsnm{{Zabalza}}, \binits{V.}},
\bauthor{\bsnm{{Astropy Contributors}}}:
\byear{2018},
\batitle{{The Astropy Project: Building an Open-science Project and Status of
  the v2.0 Core Package}}.
\bjtitle{\aj}
\bvolume{156}(\bissue{3}),
\bfpage{123}.
\doiurl{10.3847/1538-3881/aabc4f}.
\adsurl{https://ui.adsabs.harvard.edu/abs/2018AJ....156..123A}.
\end{barticle}
\endbibitem

\bibitem[\protect\citeauthoryear{{Astropy Collaboration}
  {et~al.}}{2022}]{astropy:2022}
\begin{barticle}
\bauthor{\bsnm{{Astropy Collaboration}}},
\bauthor{\bsnm{{Price-Whelan}}, \binits{A.M.}},
\bauthor{\bsnm{{Lim}}, \binits{P.L.}},
\bauthor{\bsnm{{Earl}}, \binits{N.}},
\bauthor{\bsnm{{Starkman}}, \binits{N.}},
\bauthor{\bsnm{{Bradley}}, \binits{L.}},
\bauthor{\bsnm{{Shupe}}, \binits{D.L.}},
\bauthor{\bsnm{{Patil}}, \binits{A.A.}},
\bauthor{\bsnm{{Corrales}}, \binits{L.}},
\bauthor{\bsnm{{Brasseur}}, \binits{C.E.}},
\bauthor{\bsnm{{N{"o}the}}, \binits{M.}},
\bauthor{\bsnm{{Donath}}, \binits{A.}},
\bauthor{\bsnm{{Tollerud}}, \binits{E.}},
\bauthor{\bsnm{{Morris}}, \binits{B.M.}},
\bauthor{\bsnm{{Ginsburg}}, \binits{A.}},
\bauthor{\bsnm{{Vaher}}, \binits{E.}},
\bauthor{\bsnm{{Weaver}}, \binits{B.A.}},
\bauthor{\bsnm{{Tocknell}}, \binits{J.}},
\bauthor{\bsnm{{Jamieson}}, \binits{W.}},
\bauthor{\bsnm{{van Kerkwijk}}, \binits{M.H.}},
\bauthor{\bsnm{{Robitaille}}, \binits{T.P.}},
\bauthor{\bsnm{{Merry}}, \binits{B.}},
\bauthor{\bsnm{{Bachetti}}, \binits{M.}},
\bauthor{\bsnm{{G{"u}nther}}, \binits{H.M.}},
\bauthor{\bsnm{{Aldcroft}}, \binits{T.L.}},
\bauthor{\bsnm{{Alvarado-Montes}}, \binits{J.A.}},
\bauthor{\bsnm{{Archibald}}, \binits{A.M.}},
\bauthor{\bsnm{{B{'o}di}}, \binits{A.}},
\bauthor{\bsnm{{Bapat}}, \binits{S.}},
\bauthor{\bsnm{{Barentsen}}, \binits{G.}},
\bauthor{\bsnm{{Baz{'a}n}}, \binits{J.}},
\bauthor{\bsnm{{Biswas}}, \binits{M.}},
\bauthor{\bsnm{{Boquien}}, \binits{M.}},
\bauthor{\bsnm{{Burke}}, \binits{D.J.}},
\bauthor{\bsnm{{Cara}}, \binits{D.}},
\bauthor{\bsnm{{Cara}}, \binits{M.}},
\bauthor{\bsnm{{Conroy}}, \binits{K.E.}},
\bauthor{\bsnm{{Conseil}}, \binits{S.}},
\bauthor{\bsnm{{Craig}}, \binits{M.W.}},
\bauthor{\bsnm{{Cross}}, \binits{R.M.}},
\bauthor{\bsnm{{Cruz}}, \binits{K.L.}},
\bauthor{\bsnm{{D'Eugenio}}, \binits{F.}},
\bauthor{\bsnm{{Dencheva}}, \binits{N.}},
\bauthor{\bsnm{{Devillepoix}}, \binits{H.A.R.}},
\bauthor{\bsnm{{Dietrich}}, \binits{J.P.}},
\bauthor{\bsnm{{Eigenbrot}}, \binits{A.D.}},
\bauthor{\bsnm{{Erben}}, \binits{T.}},
\bauthor{\bsnm{{Ferreira}}, \binits{L.}},
\bauthor{\bsnm{{Foreman-Mackey}}, \binits{D.}},
\bauthor{\bsnm{{Fox}}, \binits{R.}},
\bauthor{\bsnm{{Freij}}, \binits{N.}},
\bauthor{\bsnm{{Garg}}, \binits{S.}},
\bauthor{\bsnm{{Geda}}, \binits{R.}},
\bauthor{\bsnm{{Glattly}}, \binits{L.}},
\bauthor{\bsnm{{Gondhalekar}}, \binits{Y.}},
\bauthor{\bsnm{{Gordon}}, \binits{K.D.}},
\bauthor{\bsnm{{Grant}}, \binits{D.}},
\bauthor{\bsnm{{Greenfield}}, \binits{P.}},
\bauthor{\bsnm{{Groener}}, \binits{A.M.}},
\bauthor{\bsnm{{Guest}}, \binits{S.}},
\bauthor{\bsnm{{Gurovich}}, \binits{S.}},
\bauthor{\bsnm{{Handberg}}, \binits{R.}},
\bauthor{\bsnm{{Hart}}, \binits{A.}},
\bauthor{\bsnm{{Hatfield-Dodds}}, \binits{Z.}},
\bauthor{\bsnm{{Homeier}}, \binits{D.}},
\bauthor{\bsnm{{Hosseinzadeh}}, \binits{G.}},
\bauthor{\bsnm{{Jenness}}, \binits{T.}},
\bauthor{\bsnm{{Jones}}, \binits{C.K.}},
\bauthor{\bsnm{{Joseph}}, \binits{P.}},
\bauthor{\bsnm{{Kalmbach}}, \binits{J.B.}},
\bauthor{\bsnm{{Karamehmetoglu}}, \binits{E.}},
\bauthor{\bsnm{{Ka{l}uszy{'n}ski}}, \binits{M.}},
\bauthor{\bsnm{{Kelley}}, \binits{M.S.P.}},
\bauthor{\bsnm{{Kern}}, \binits{N.}},
\bauthor{\bsnm{{Kerzendorf}}, \binits{W.E.}},
\bauthor{\bsnm{{Koch}}, \binits{E.W.}},
\bauthor{\bsnm{{Kulumani}}, \binits{S.}},
\bauthor{\bsnm{{Lee}}, \binits{A.}},
\bauthor{\bsnm{{Ly}}, \binits{C.}},
\bauthor{\bsnm{{Ma}}, \binits{Z.}},
\bauthor{\bsnm{{MacBride}}, \binits{C.}},
\bauthor{\bsnm{{Maljaars}}, \binits{J.M.}},
\bauthor{\bsnm{{Muna}}, \binits{D.}},
\bauthor{\bsnm{{Murphy}}, \binits{N.A.}},
\bauthor{\bsnm{{Norman}}, \binits{H.}},
\bauthor{\bsnm{{O'Steen}}, \binits{R.}},
\bauthor{\bsnm{{Oman}}, \binits{K.A.}},
\bauthor{\bsnm{{Pacifici}}, \binits{C.}},
\bauthor{\bsnm{{Pascual}}, \binits{S.}},
\bauthor{\bsnm{{Pascual-Granado}}, \binits{J.}},
\bauthor{\bsnm{{Patil}}, \binits{R.R.}},
\bauthor{\bsnm{{Perren}}, \binits{G.I.}},
\bauthor{\bsnm{{Pickering}}, \binits{T.E.}},
\bauthor{\bsnm{{Rastogi}}, \binits{T.}},
\bauthor{\bsnm{{Roulston}}, \binits{B.R.}},
\bauthor{\bsnm{{Ryan}}, \binits{D.F.}},
\bauthor{\bsnm{{Rykoff}}, \binits{E.S.}},
\bauthor{\bsnm{{Sabater}}, \binits{J.}},
\bauthor{\bsnm{{Sakurikar}}, \binits{P.}},
\bauthor{\bsnm{{Salgado}}, \binits{J.}},
\bauthor{\bsnm{{Sanghi}}, \binits{A.}},
\bauthor{\bsnm{{Saunders}}, \binits{N.}},
\bauthor{\bsnm{{Savchenko}}, \binits{V.}},
\bauthor{\bsnm{{Schwardt}}, \binits{L.}},
\bauthor{\bsnm{{Seifert-Eckert}}, \binits{M.}},
\bauthor{\bsnm{{Shih}}, \binits{A.Y.}},
\bauthor{\bsnm{{Jain}}, \binits{A.S.}},
\bauthor{\bsnm{{Shukla}}, \binits{G.}},
\bauthor{\bsnm{{Sick}}, \binits{J.}},
\bauthor{\bsnm{{Simpson}}, \binits{C.}},
\bauthor{\bsnm{{Singanamalla}}, \binits{S.}},
\bauthor{\bsnm{{Singer}}, \binits{L.P.}},
\bauthor{\bsnm{{Singhal}}, \binits{J.}},
\bauthor{\bsnm{{Sinha}}, \binits{M.}},
\bauthor{\bsnm{{Sip{H{o}}cz}}, \binits{B.M.}},
\bauthor{\bsnm{{Spitler}}, \binits{L.R.}},
\bauthor{\bsnm{{Stansby}}, \binits{D.}},
\bauthor{\bsnm{{Streicher}}, \binits{O.}},
\bauthor{\bsnm{{{{S}}umak}}, \binits{J.}},
\bauthor{\bsnm{{Swinbank}}, \binits{J.D.}},
\bauthor{\bsnm{{Taranu}}, \binits{D.S.}},
\bauthor{\bsnm{{Tewary}}, \binits{N.}},
\bauthor{\bsnm{{Tremblay}}, \binits{G.R.}},
\bauthor{\bsnm{{Val-Borro}}, \binits{M.d.}},
\bauthor{\bsnm{{Van Kooten}}, \binits{S.J.}},
\bauthor{\bsnm{{Vasovi{'c}}}, \binits{Z.}},
\bauthor{\bsnm{{Verma}}, \binits{S.}},
\bauthor{\bsnm{{de Miranda Cardoso}}, \binits{J.V.}},
\bauthor{\bsnm{{Williams}}, \binits{P.K.G.}},
\bauthor{\bsnm{{Wilson}}, \binits{T.J.}},
\bauthor{\bsnm{{Winkel}}, \binits{B.}},
\bauthor{\bsnm{{Wood-Vasey}}, \binits{W.M.}},
\bauthor{\bsnm{{Xue}}, \binits{R.}},
\bauthor{\bsnm{{Yoachim}}, \binits{P.}},
\bauthor{\bsnm{{Zhang}}, \binits{C.}},
\bauthor{\bsnm{{Zonca}}, \binits{A.}},
\bauthor{\bsnm{{Astropy Project Contributors}}}:
\byear{2022},
\batitle{{The Astropy Project: Sustaining and Growing a Community-oriented
  Open-source Project and the Latest Major Release (v5.0) of the Core
  Package}}.
\bjtitle{\apj}
\bvolume{935}(\bissue{2}),
\bfpage{167}.
\doiurl{10.3847/1538-4357/ac7c74}.
\adsurl{https://ui.adsabs.harvard.edu/abs/2022ApJ...935..167A}.
\end{barticle}
\endbibitem

\bibitem[\protect\citeauthoryear{{Avignon} {et~al.}}{1989}]{avignon1989}
\begin{barticle}
\bauthor{\bsnm{{Avignon}}, \binits{Y.}},
\bauthor{\bsnm{{Bonmartin}}, \binits{J.}},
\bauthor{\bsnm{{Bouteille}}, \binits{A.}},
\bauthor{\bsnm{{Clavelier}}, \binits{B.}},
\bauthor{\bsnm{{Hulot}}, \binits{E.}},
\bauthor{\bsnm{{Issartel}}, \binits{M.P.}},
\bauthor{\bsnm{{Kerdraon}}, \binits{A.}},
\bauthor{\bsnm{{Klein}}, \binits{K.-L.}},
\bauthor{\bsnm{{Lantos}}, \binits{P.}},
\bauthor{\bsnm{{Mercier}}, \binits{C.}},
\bauthor{\bsnm{{Pick}}, \binits{M.}},
\bauthor{\bsnm{{Raoult}}, \binits{A.}},
\bauthor{\bsnm{{Rigaud}}, \binits{D.}},
\bauthor{\bsnm{{Trottet}}, \binits{G.}},
\bauthor{\bsnm{{Vilmer}}, \binits{N.}},
\bauthor{\bsnm{{Chantelat}}, \binits{C.}},
\bauthor{\bsnm{{Chapuis}}, \binits{M.}},
\bauthor{\bsnm{{Chapuis}}, \binits{Y.}},
\bauthor{\bsnm{{Coffre}}, \binits{A.}},
\bauthor{\bsnm{{Couteret}}, \binits{C.}},
\bauthor{\bsnm{{Darchy}}, \binits{B.}},
\bauthor{\bsnm{{Gu{\'e}niau}}, \binits{P.}},
\bauthor{\bsnm{{Lalardie}}, \binits{D.}},
\bauthor{\bsnm{{Picard}}, \binits{P.}},
\bauthor{\bsnm{{Tocqueville}}, \binits{R.}}:
\byear{1989},
\batitle{{The Mark IV Nan{\c{c}}ay Radioheliograph.}}
\bjtitle{\solphys}
\bvolume{120}(\bissue{1}),
\bfpage{193}.
\adsurl{https://ui.adsabs.harvard.edu/abs/1989SoPh..120..193A}.
\end{barticle}
\endbibitem

\bibitem[\protect\citeauthoryear{{Bastian}}{1994}]{bastian1994}
\begin{barticle}
\bauthor{\bsnm{{Bastian}}, \binits{T.S.}}:
\byear{1994},
\batitle{{Angular Scattering of Solar Radio Emission by Coronal Turbulence}}.
\bjtitle{\apj}
\bvolume{426},
\bfpage{774}.
\doiurl{10.1086/174114}.
\adsurl{https://ui.adsabs.harvard.edu/abs/1994ApJ...426..774B}.
\end{barticle}
\endbibitem

\bibitem[\protect\citeauthoryear{{Bonmartin} {et~al.}}{1983}]{bonmartin1983}
\begin{barticle}
\bauthor{\bsnm{{Bonmartin}}, \binits{J.}},
\bauthor{\bsnm{{Bouteille}}, \binits{A.}},
\bauthor{\bsnm{{Clavelier}}, \binits{B.}},
\bauthor{\bsnm{{Issartel}}, \binits{M.P.}},
\bauthor{\bsnm{{Kerdraon}}, \binits{A.}},
\bauthor{\bsnm{{Lantos}}, \binits{M.F.}},
\bauthor{\bsnm{{Lantos}}, \binits{P.}},
\bauthor{\bsnm{{Mercier}}, \binits{C.}},
\bauthor{\bsnm{{Pick}}, \binits{M.}},
\bauthor{\bsnm{{Raoult}}, \binits{A.}},
\bauthor{\bsnm{{Trottet}}, \binits{G.}},
\bauthor{\bsnm{{Bruley}}, \binits{M.}},
\bauthor{\bsnm{{Chantelat}}, \binits{C.}},
\bauthor{\bsnm{{Chapuis}}, \binits{M.}},
\bauthor{\bsnm{{Couteret}}, \binits{C.}},
\bauthor{\bsnm{{Gueniau}}, \binits{P.}},
\bauthor{\bsnm{{Lalardie}}, \binits{D.}},
\bauthor{\bsnm{{Picard}}, \binits{R.P.}},
\bauthor{\bsnm{{Tocqueville}}, \binits{B.}},
\bauthor{\bsnm{{Henry}}, \binits{J.C.}},
\bauthor{\bsnm{{Renaud}}, \binits{J.}}:
\byear{1983},
\batitle{{The mark III Nan{\c{c}}ay radioheliograph The Radioheliograph
  Group}}.
\bjtitle{\solphys}
\bvolume{88}(\bissue{1-2}),
\bfpage{383}.
\doiurl{10.1007/BF00196201}.
\adsurl{https://ui.adsabs.harvard.edu/abs/1983SoPh...88..383B}.
\end{barticle}
\endbibitem

\bibitem[\protect\citeauthoryear{{Briggs}}{1995}]{briggs1995}
\begin{bchapter}
\bauthor{\bsnm{{Briggs}}, \binits{D.S.}}:
\byear{1995},
\bctitle{{High Fidelity Interferometric Imaging: Robust Weighting and NNLS
  Deconvolution}}.
In: \bbtitle{American Astronomical Society Meeting Abstracts},
\bsertitle{American Astronomical Society Meeting Abstracts}
\bseriesno{187},
\bfpage{112.02}.
\adsurl{https://ui.adsabs.harvard.edu/abs/1995AAS...18711202B}.
\end{bchapter}
\endbibitem

\bibitem[\protect\citeauthoryear{Buch {et~al.}}{2022}]{Buch2022}
\begin{barticle}
\bauthor{\bsnm{Buch}, \binits{K.D.}},
\bauthor{\bsnm{Kale}, \binits{R.}},
\bauthor{\bsnm{Naik}, \binits{K.D.}},
\bauthor{\bsnm{Aragade}, \binits{R.}},
\bauthor{\bsnm{Muley}, \binits{M.}},
\bauthor{\bsnm{Kudale}, \binits{S.}},
\bauthor{\bsnm{Ajith~Kumar}, \binits{B.}}:
\byear{2022},
\batitle{Performance analysis techniques for real-time broadband rfi filtering
  system of ugmrt}.
\bjtitle{Journal of Astronomical Instrumentation}
\bvolume{11}(\bissue{02}),
\bfpage{2250008}.
\doiurl{10.1142/S2251171722500088}.
\burl{https://doi.org/10.1142/S2251171722500088}.
\end{barticle}
\endbibitem

\bibitem[\protect\citeauthoryear{{Buch} {et~al.}}{2023}]{Buch2023}
\begin{barticle}
\bauthor{\bsnm{{Buch}}, \binits{K.D.}},
\bauthor{\bsnm{{Kale}}, \binits{R.}},
\bauthor{\bsnm{{Muley}}, \binits{M.}},
\bauthor{\bsnm{{Kudale}}, \binits{S.}},
\bauthor{\bsnm{{Ajithkumar}}, \binits{B.}}:
\byear{2023},
\batitle{{Real-time RFI filtering for uGMRT: Overview of the released system
  and relevance to the SKA}}.
\bjtitle{Journal of Astrophysics and Astronomy}
\bvolume{44}(\bissue{1}),
\bfpage{37}.
\doiurl{10.1007/s12036-023-09919-x}.
\adsurl{https://ui.adsabs.harvard.edu/abs/2023JApA...44...37B}.
\end{barticle}
\endbibitem

\bibitem[\protect\citeauthoryear{{Elgar{\o}y}}{1977}]{elgaroy1977}
\begin{bbook}
\bauthor{\bsnm{{Elgar{\o}y}}, \binits{E.{\O}.}}:
\byear{1977},
\bbtitle{{Solar noise storms.}}
\adsurl{https://ui.adsabs.harvard.edu/abs/1977sns..book.....E}.
\end{bbook}
\endbibitem

\bibitem[\protect\citeauthoryear{{Gupta} {et~al.}}{2017}]{gupta2017}
\begin{barticle}
\bauthor{\bsnm{{Gupta}}, \binits{Y.}},
\bauthor{\bsnm{{Ajithkumar}}, \binits{B.}},
\bauthor{\bsnm{{Kale}}, \binits{H.S.}},
\bauthor{\bsnm{{Nayak}}, \binits{S.}},
\bauthor{\bsnm{{Sabhapathy}}, \binits{S.}},
\bauthor{\bsnm{{Sureshkumar}}, \binits{S.}},
\bauthor{\bsnm{{Swami}}, \binits{R.V.}},
\bauthor{\bsnm{{Chengalur}}, \binits{J.N.}},
\bauthor{\bsnm{{Ghosh}}, \binits{S.K.}},
\bauthor{\bsnm{{Ishwara-Chandra}}, \binits{C.H.}},
\bauthor{\bsnm{{Joshi}}, \binits{B.C.}},
\bauthor{\bsnm{{Kanekar}}, \binits{N.}},
\bauthor{\bsnm{{Lal}}, \binits{D.V.}},
\bauthor{\bsnm{{Roy}}, \binits{S.}}:
\byear{2017},
\batitle{{The upgraded GMRT: opening new windows on the radio Universe}}.
\bjtitle{Current Science}
\bvolume{113}(\bissue{4}),
\bfpage{707}.
\doiurl{10.18520/cs/v113/i04/707-714}.
\adsurl{https://ui.adsabs.harvard.edu/abs/2017CSci..113..707G}.
\end{barticle}
\endbibitem

\bibitem[\protect\citeauthoryear{{Habbal}, {Ellman}, and
  {Gonzalez}}{1989}]{habbal1989}
\begin{barticle}
\bauthor{\bsnm{{Habbal}}, \binits{S.R.}},
\bauthor{\bsnm{{Ellman}}, \binits{N.E.}},
\bauthor{\bsnm{{Gonzalez}}, \binits{R.}}:
\byear{1989},
\batitle{{Synthesis Mapping of a Solar Type I Storm Simultaneously at 90 and 20
  Centimeters with the VLA}}.
\bjtitle{\apj}
\bvolume{342},
\bfpage{594}.
\doiurl{10.1086/167619}.
\adsurl{https://ui.adsabs.harvard.edu/abs/1989ApJ...342..594H}.
\end{barticle}
\endbibitem

\bibitem[\protect\citeauthoryear{Harris {et~al.}}{2020}]{Harris2020}
\begin{barticle}
\bauthor{\bsnm{Harris}, \binits{C.R.}},
\bauthor{\bsnm{Millman}, \binits{K.J.}},
\bauthor{\bparticle{van~der} \bsnm{Walt}, \binits{S.J.}},
\bauthor{\bsnm{Gommers}, \binits{R.}},
\bauthor{\bsnm{Virtanen}, \binits{P.}},
\bauthor{\bsnm{Cournapeau}, \binits{D.}},
\bauthor{\bsnm{Wieser}, \binits{E.}},
\bauthor{\bsnm{Taylor}, \binits{J.}},
\bauthor{\bsnm{Berg}, \binits{S.}},
\bauthor{\bsnm{Smith}, \binits{N.J.}},
\bauthor{\bsnm{Kern}, \binits{R.}},
\bauthor{\bsnm{Picus}, \binits{M.}},
\bauthor{\bsnm{Hoyer}, \binits{S.}},
\bauthor{\bparticle{van} \bsnm{Kerkwijk}, \binits{M.H.}},
\bauthor{\bsnm{Brett}, \binits{M.}},
\bauthor{\bsnm{Haldane}, \binits{A.}},
\bauthor{\bparticle{del} \bsnm{R{\'i}o}, \binits{J.F.}},
\bauthor{\bsnm{Wiebe}, \binits{M.}},
\bauthor{\bsnm{Peterson}, \binits{P.}},
\bauthor{\bsnm{G{\'e}rard-Marchant}, \binits{P.}},
\bauthor{\bsnm{Sheppard}, \binits{K.}},
\bauthor{\bsnm{Reddy}, \binits{T.}},
\bauthor{\bsnm{Weckesser}, \binits{W.}},
\bauthor{\bsnm{Abbasi}, \binits{H.}},
\bauthor{\bsnm{Gohlke}, \binits{C.}},
\bauthor{\bsnm{Oliphant}, \binits{T.E.}}:
\byear{2020},
\batitle{Array programming with numpy}.
\bjtitle{Nature}
\bvolume{585}(\bissue{7825}),
\bfpage{357}.
\doiurl{10.1038/s41586-020-2649-2}.
\burl{https://doi.org/10.1038/s41586-020-2649-2}.
\end{barticle}
\endbibitem

\bibitem[\protect\citeauthoryear{Hunter}{2007}]{Hunter:2007}
\begin{barticle}
\bauthor{\bsnm{Hunter}, \binits{J.D.}}:
\byear{2007},
\batitle{Matplotlib: A 2d graphics environment}.
\bjtitle{Computing in Science \& Engineering}
\bvolume{9}(\bissue{3}),
\bfpage{90}.
\doiurl{10.1109/MCSE.2007.55}.
\end{barticle}
\endbibitem

\bibitem[\protect\citeauthoryear{{Kerdraon} and {Delouis}}{1997}]{kerdraon1997}
\begin{bchapter}
\bauthor{\bsnm{{Kerdraon}}, \binits{A.}},
\bauthor{\bsnm{{Delouis}}, \binits{J.-M.}}:
\byear{1997},
\bctitle{{The Nan{\c{c}}ay Radioheliograph}}.
In: \beditor{\bsnm{{Trottet}}, \binits{G.}} (ed.)
\bbtitle{Coronal Physics from Radio and Space Observations}
\bseriesno{483},
\bfpage{192}.
\doiurl{10.1007/BFb0106458}.
\adsurl{https://ui.adsabs.harvard.edu/abs/1997LNP...483..192K}.
\end{bchapter}
\endbibitem

\bibitem[\protect\citeauthoryear{{Kerdraon} {et~al.}}{1988}]{kerdraon1988}
\begin{barticle}
\bauthor{\bsnm{{Kerdraon}}, \binits{A.}},
\bauthor{\bsnm{{Lang}}, \binits{K.R.}},
\bauthor{\bsnm{{Trottet}}, \binits{G.}},
\bauthor{\bsnm{{Willson}}, \binits{R.F.}}:
\byear{1988},
\batitle{{High resolution VLA-Nancay observations of the sun}}.
\bjtitle{Advances in Space Research}
\bvolume{8}(\bissue{11}),
\bfpage{45}.
\doiurl{10.1016/0273-1177(88)90294-3}.
\adsurl{https://ui.adsabs.harvard.edu/abs/1988AdSpR...8k..45K}.
\end{barticle}
\endbibitem

\bibitem[\protect\citeauthoryear{{Kontar} {et~al.}}{2019}]{kontar2019}
\begin{barticle}
\bauthor{\bsnm{{Kontar}}, \binits{E.P.}},
\bauthor{\bsnm{{Chen}}, \binits{X.}},
\bauthor{\bsnm{{Chrysaphi}}, \binits{N.}},
\bauthor{\bsnm{{Jeffrey}}, \binits{N.L.S.}},
\bauthor{\bsnm{{Emslie}}, \binits{A.G.}},
\bauthor{\bsnm{{Krupar}}, \binits{V.}},
\bauthor{\bsnm{{Maksimovic}}, \binits{M.}},
\bauthor{\bsnm{{Gordovskyy}}, \binits{M.}},
\bauthor{\bsnm{{Browning}}, \binits{P.K.}}:
\byear{2019},
\batitle{{Anisotropic Radio-wave Scattering and the Interpretation of Solar
  Radio Emission Observations}}.
\bjtitle{\apj}
\bvolume{884}(\bissue{2}),
\bfpage{122}.
\doiurl{10.3847/1538-4357/ab40bb}.
\adsurl{https://ui.adsabs.harvard.edu/abs/2019ApJ...884..122K}.
\end{barticle}
\endbibitem

\bibitem[\protect\citeauthoryear{Kontar {et~al.}}{2023}]{kontar2023anisotropic}
\begin{barticle}
\bauthor{\bsnm{Kontar}, \binits{E.P.}},
\bauthor{\bsnm{Emslie}, \binits{A.G.}},
\bauthor{\bsnm{Clarkson}, \binits{D.L.}},
\bauthor{\bsnm{Chen}, \binits{X.}},
\bauthor{\bsnm{Chrysaphi}, \binits{N.}},
\bauthor{\bsnm{Azzollini}, \binits{F.}},
\bauthor{\bsnm{Jeffrey}, \binits{N.L.}},
\bauthor{\bsnm{Gordovskyy}, \binits{M.}}:
\byear{2023},
\batitle{An anisotropic density turbulence model from the sun to 1 au derived
  from radio observations}.
\bjtitle{The Astrophysical Journal}
\bvolume{956}(\bissue{2}),
\bfpage{112}.
\end{barticle}
\endbibitem

\bibitem[\protect\citeauthoryear{{Lang} and {Willson}}{1987}]{lang1987}
\begin{barticle}
\bauthor{\bsnm{{Lang}}, \binits{K.R.}},
\bauthor{\bsnm{{Willson}}, \binits{R.F.}}:
\byear{1987},
\batitle{{VLA Observations of a Solar Noise Storm}}.
\bjtitle{\apj}
\bvolume{319},
\bfpage{514}.
\doiurl{10.1086/165474}.
\adsurl{https://ui.adsabs.harvard.edu/abs/1987ApJ...319..514L}.
\end{barticle}
\endbibitem

\bibitem[\protect\citeauthoryear{{Lemen} {et~al.}}{2012}]{lemen2012}
\begin{barticle}
\bauthor{\bsnm{{Lemen}}, \binits{J.R.}},
\bauthor{\bsnm{{Title}}, \binits{A.M.}},
\bauthor{\bsnm{{Akin}}, \binits{D.J.}},
\bauthor{\bsnm{{Boerner}}, \binits{P.F.}},
\bauthor{\bsnm{{Chou}}, \binits{C.}},
\bauthor{\bsnm{{Drake}}, \binits{J.F.}},
\bauthor{\bsnm{{Duncan}}, \binits{D.W.}},
\bauthor{\bsnm{{Edwards}}, \binits{C.G.}},
\bauthor{\bsnm{{Friedlaender}}, \binits{F.M.}},
\bauthor{\bsnm{{Heyman}}, \binits{G.F.}},
\bauthor{\bsnm{{Hurlburt}}, \binits{N.E.}},
\bauthor{\bsnm{{Katz}}, \binits{N.L.}},
\bauthor{\bsnm{{Kushner}}, \binits{G.D.}},
\bauthor{\bsnm{{Levay}}, \binits{M.}},
\bauthor{\bsnm{{Lindgren}}, \binits{R.W.}},
\bauthor{\bsnm{{Mathur}}, \binits{D.P.}},
\bauthor{\bsnm{{McFeaters}}, \binits{E.L.}},
\bauthor{\bsnm{{Mitchell}}, \binits{S.}},
\bauthor{\bsnm{{Rehse}}, \binits{R.A.}},
\bauthor{\bsnm{{Schrijver}}, \binits{C.J.}},
\bauthor{\bsnm{{Springer}}, \binits{L.A.}},
\bauthor{\bsnm{{Stern}}, \binits{R.A.}},
\bauthor{\bsnm{{Tarbell}}, \binits{T.D.}},
\bauthor{\bsnm{{Wuelser}}, \binits{J.-P.}},
\bauthor{\bsnm{{Wolfson}}, \binits{C.J.}},
\bauthor{\bsnm{{Yanari}}, \binits{C.}},
\bauthor{\bsnm{{Bookbinder}}, \binits{J.A.}},
\bauthor{\bsnm{{Cheimets}}, \binits{P.N.}},
\bauthor{\bsnm{{Caldwell}}, \binits{D.}},
\bauthor{\bsnm{{Deluca}}, \binits{E.E.}},
\bauthor{\bsnm{{Gates}}, \binits{R.}},
\bauthor{\bsnm{{Golub}}, \binits{L.}},
\bauthor{\bsnm{{Park}}, \binits{S.}},
\bauthor{\bsnm{{Podgorski}}, \binits{W.A.}},
\bauthor{\bsnm{{Bush}}, \binits{R.I.}},
\bauthor{\bsnm{{Scherrer}}, \binits{P.H.}},
\bauthor{\bsnm{{Gummin}}, \binits{M.A.}},
\bauthor{\bsnm{{Smith}}, \binits{P.}},
\bauthor{\bsnm{{Auker}}, \binits{G.}},
\bauthor{\bsnm{{Jerram}}, \binits{P.}},
\bauthor{\bsnm{{Pool}}, \binits{P.}},
\bauthor{\bsnm{{Soufli}}, \binits{R.}},
\bauthor{\bsnm{{Windt}}, \binits{D.L.}},
\bauthor{\bsnm{{Beardsley}}, \binits{S.}},
\bauthor{\bsnm{{Clapp}}, \binits{M.}},
\bauthor{\bsnm{{Lang}}, \binits{J.}},
\bauthor{\bsnm{{Waltham}}, \binits{N.}}:
\byear{2012},
\batitle{{The Atmospheric Imaging Assembly (AIA) on the Solar Dynamics
  Observatory (SDO)}}.
\bjtitle{\solphys}
\bvolume{275}(\bissue{1-2}),
\bfpage{17}.
\doiurl{10.1007/s11207-011-9776-8}.
\adsurl{https://ui.adsabs.harvard.edu/abs/2012SoPh..275...17L}.
\end{barticle}
\endbibitem

\bibitem[\protect\citeauthoryear{{McCauley} {et~al.}}{2019}]{mccauley2019}
\begin{barticle}
\bauthor{\bsnm{{McCauley}}, \binits{P.I.}},
\bauthor{\bsnm{{Cairns}}, \binits{I.H.}},
\bauthor{\bsnm{{White}}, \binits{S.M.}},
\bauthor{\bsnm{{Mondal}}, \binits{S.}},
\bauthor{\bsnm{{Lenc}}, \binits{E.}},
\bauthor{\bsnm{{Morgan}}, \binits{J.}},
\bauthor{\bsnm{{Oberoi}}, \binits{D.}}:
\byear{2019},
\batitle{{The Low-Frequency Solar Corona in Circular Polarization}}.
\bjtitle{\solphys}
\bvolume{294}(\bissue{8}),
\bfpage{106}.
\doiurl{10.1007/s11207-019-1502-y}.
\adsurl{https://ui.adsabs.harvard.edu/abs/2019SoPh..294..106M}.
\end{barticle}
\endbibitem

\bibitem[\protect\citeauthoryear{McLean}{1967}]{mclean1967}
\begin{barticle}
\bauthor{\bsnm{McLean}, \binits{D.}}:
\byear{1967},
\batitle{Band splitting in type ii solar radio bursts}.
\bjtitle{Publications of the Astronomical Society of Australia}
\bvolume{1}(\bissue{2}),
\bfpage{47}.
\end{barticle}
\endbibitem

\bibitem[\protect\citeauthoryear{{Melrose}}{1980}]{melrose1980}
\begin{barticle}
\bauthor{\bsnm{{Melrose}}, \binits{D.B.}}:
\byear{1980},
\batitle{{A Plasma Emission Mechanism for Type-I Solar Radio Emission}}.
\bjtitle{\solphys}
\bvolume{67}(\bissue{2}),
\bfpage{357}.
\doiurl{10.1007/BF00149813}.
\adsurl{https://ui.adsabs.harvard.edu/abs/1980SoPh...67..357M}.
\end{barticle}
\endbibitem

\bibitem[\protect\citeauthoryear{{Mercier} {et~al.}}{2006}]{mercier2006}
\begin{barticle}
\bauthor{\bsnm{{Mercier}}, \binits{C.}},
\bauthor{\bsnm{{Subramanian}}, \binits{P.}},
\bauthor{\bsnm{{Kerdraon}}, \binits{A.}},
\bauthor{\bsnm{{Pick}}, \binits{M.}},
\bauthor{\bsnm{{Ananthakrishnan}}, \binits{S.}},
\bauthor{\bsnm{{Janardhan}}, \binits{P.}}:
\byear{2006},
\batitle{{Combining visibilities from the giant meterwave radio telescope and
  the Nancay radio heliograph. High dynamic range snapshot images of the solar
  corona at 327 MHz}}.
\bjtitle{\aap}
\bvolume{447}(\bissue{3}),
\bfpage{1189}.
\doiurl{10.1051/0004-6361:20053621}.
\adsurl{https://ui.adsabs.harvard.edu/abs/2006A&A...447.1189M}.
\end{barticle}
\endbibitem

\bibitem[\protect\citeauthoryear{{Mercier} {et~al.}}{2015}]{mercier2015}
\begin{barticle}
\bauthor{\bsnm{{Mercier}}, \binits{C.}},
\bauthor{\bsnm{{Subramanian}}, \binits{P.}},
\bauthor{\bsnm{{Chambe}}, \binits{G.}},
\bauthor{\bsnm{{Janardhan}}, \binits{P.}}:
\byear{2015},
\batitle{{The structure of solar radio noise storms}}.
\bjtitle{\aap}
\bvolume{576},
\bfpage{A136}.
\doiurl{10.1051/0004-6361/201321064}.
\adsurl{https://ui.adsabs.harvard.edu/abs/2015A&A...576A.136M}.
\end{barticle}
\endbibitem

\bibitem[\protect\citeauthoryear{{Mohan}}{2021}]{mohan2021}
\begin{barticle}
\bauthor{\bsnm{{Mohan}}, \binits{A.}}:
\byear{2021},
\batitle{{Characterising coronal turbulence using snapshot imaging of radio
  bursts in 80-200 MHz}}.
\bjtitle{\aap}
\bvolume{655},
\bfpage{A77}.
\doiurl{10.1051/0004-6361/202142029}.
\adsurl{https://ui.adsabs.harvard.edu/abs/2021A&A...655A..77M}.
\end{barticle}
\endbibitem

\bibitem[\protect\citeauthoryear{{Mohan} {et~al.}}{2019}]{mohan2019b}
\begin{barticle}
\bauthor{\bsnm{{Mohan}}, \binits{A.}},
\bauthor{\bsnm{{McCauley}}, \binits{P.I.}},
\bauthor{\bsnm{{Oberoi}}, \binits{D.}},
\bauthor{\bsnm{{Mastrano}}, \binits{A.}}:
\byear{2019},
\batitle{{A Weak Coronal Heating Event Associated with Periodic Particle
  Acceleration Episodes}}.
\bjtitle{Astrophysical Journal}
\bvolume{883}(\bissue{1}),
\bfpage{45}.
\doiurl{10.3847/1538-4357/ab3a94}.
\adsurl{https://ui.adsabs.harvard.edu/abs/2019ApJ...883...45M}.
\end{barticle}
\endbibitem

\bibitem[\protect\citeauthoryear{{Mondal}
  {et~al.}}{2024}]{mondal2024_noise_storm}
\begin{barticle}
\bauthor{\bsnm{{Mondal}}, \binits{S.}},
\bauthor{\bsnm{{Kansabanik}}, \binits{D.}},
\bauthor{\bsnm{{Oberoi}}, \binits{D.}},
\bauthor{\bsnm{{Dey}}, \binits{S.}}:
\byear{2024},
\batitle{{New Insights into Type-I Solar Noise Storms from High Angular
  Resolution Spectroscopic Imaging with the Upgraded Giant Metrewave Radio
  Telescope}}.
\bjtitle{\apj}
\bvolume{975}(\bissue{1}),
\bfpage{122}.
\doiurl{10.3847/1538-4357/ad77c2}.
\adsurl{https://ui.adsabs.harvard.edu/abs/2024ApJ...975..122M}.
\end{barticle}
\endbibitem

\bibitem[\protect\citeauthoryear{{Mugundhan} {et~al.}}{2018}]{mugundhan2018}
\begin{barticle}
\bauthor{\bsnm{{Mugundhan}}, \binits{V.}},
\bauthor{\bsnm{{Ramesh}}, \binits{R.}},
\bauthor{\bsnm{{Kathiravan}}, \binits{C.}},
\bauthor{\bsnm{{Gireesh}}, \binits{G.V.S.}},
\bauthor{\bsnm{{Kumari}}, \binits{A.}},
\bauthor{\bsnm{{Hariharan}}, \binits{K.}},
\bauthor{\bsnm{{Barve}}, \binits{I.V.}}:
\byear{2018},
\batitle{{The First Low-frequency Radio Observations of the Solar Corona on
  {\ensuremath{\approx}}200 km Long Interferometer Baseline}}.
\bjtitle{\apjl}
\bvolume{855}(\bissue{1}),
\bfpage{L8}.
\doiurl{10.3847/2041-8213/aaaf64}.
\adsurl{https://ui.adsabs.harvard.edu/abs/2018ApJ...855L...8M}.
\end{barticle}
\endbibitem

\bibitem[\protect\citeauthoryear{Mumford {et~al.}}{2021}]{sunpy2.1.0}
\begin{botherref}
\oauthor{\bsnm{Mumford}, \binits{S.J.}},
\oauthor{\bsnm{Freij}, \binits{N.}},
\oauthor{\bsnm{Christe}, \binits{S.}},
\oauthor{\bsnm{Ireland}, \binits{J.}},
\oauthor{\bsnm{Mayer}, \binits{F.}},
\oauthor{\bsnm{Hughitt}, \binits{V.K.}},
\oauthor{\bsnm{Shih}, \binits{A.Y.}},
\oauthor{\bsnm{Ryan}, \binits{D.F.}},
\oauthor{\bsnm{Liedtke}, \binits{S.}},
\oauthor{\bsnm{Stansby}, \binits{D.}},
\oauthor{\bsnm{Pérez-Suárez}, \binits{D.}},
\oauthor{\bsnm{I.}, \binits{V.K.}},
\oauthor{\bsnm{Chakraborty}, \binits{P.}},
\oauthor{\bsnm{Inglis}, \binits{A.}},
\oauthor{\bsnm{Pattnaik}, \binits{P.}},
\oauthor{\bsnm{Sipőcz}, \binits{B.}},
\oauthor{\bsnm{Hayes}, \binits{L.}},
\oauthor{\bsnm{Sharma}, \binits{R.}},
\oauthor{\bsnm{Leonard}, \binits{A.}},
\oauthor{\bsnm{Hewett}, \binits{R.}},
\oauthor{\bsnm{Hamilton}, \binits{A.}},
\oauthor{\bsnm{Panda}, \binits{A.}},
\oauthor{\bsnm{Earnshaw}, \binits{M.}},
\oauthor{\bsnm{Choudhary}, \binits{N.}},
\oauthor{\bsnm{Kumar}, \binits{A.}},
\oauthor{\bsnm{Singh}, \binits{R.}},
\oauthor{\bsnm{Barnes}, \binits{W.}},
\oauthor{\bsnm{Chanda}, \binits{P.}},
\oauthor{\bsnm{Haque}, \binits{M.A.}},
\oauthor{\bsnm{Kirk}, \binits{M.S.}},
\oauthor{\bsnm{Konge}, \binits{S.}},
\oauthor{\bsnm{Mueller}, \binits{M.}},
\oauthor{\bsnm{Srivastava}, \binits{R.}},
\oauthor{\bsnm{Manhas}, \binits{A.}},
\oauthor{\bsnm{Jain}, \binits{Y.}},
\oauthor{\bsnm{Bennett}, \binits{S.}},
\oauthor{\bsnm{Baruah}, \binits{A.}},
\oauthor{\bsnm{Arbolante}, \binits{Q.}},
\oauthor{\bsnm{Charlton}, \binits{M.}},
\oauthor{\bsnm{Maloney}, \binits{S.}},
\oauthor{\bsnm{Mishra}, \binits{S.}},
\oauthor{\bsnm{Chorley}, \binits{N.}},
\oauthor{\bsnm{Himanshu}},
\oauthor{\bsnm{Modi}, \binits{S.}},
\oauthor{\bsnm{Mason}, \binits{J.P.}},
\oauthor{\bsnm{Sharma}, \binits{Y.}},
\oauthor{\bsnm{Naman9639}},
\oauthor{\bsnm{Bobra}, \binits{M.G.}},
\oauthor{\bsnm{Rozo}, \binits{J.I.C.}},
\oauthor{\bsnm{Manley}, \binits{L.}},
\oauthor{\bsnm{Chatterjee}, \binits{A.}},
\oauthor{\bsnm{Bazán}, \binits{J.}},
\oauthor{\bsnm{Jain}, \binits{S.}},
\oauthor{\bsnm{Evans}, \binits{J.}},
\oauthor{\bsnm{Ghosh}, \binits{S.}},
\oauthor{\bsnm{Malocha}, \binits{M.}},
\oauthor{\bsnm{Visscher}, \binits{R.D.}},
\oauthor{\bsnm{Singh}, \binits{R.R.}},
\oauthor{\bsnm{Stańczak}, \binits{D.}},
\oauthor{\bsnm{Verma}, \binits{S.}},
\oauthor{\bsnm{Airmansmith97}},
\oauthor{\bsnm{Agrawal}, \binits{A.}},
\oauthor{\bsnm{Buddhika}, \binits{D.}},
\oauthor{\bsnm{Pathak}, \binits{H.}},
\oauthor{\bsnm{Sharma}, \binits{S.}},
\oauthor{\bsnm{Alam}, \binits{A.}},
\oauthor{\bsnm{Bates}, \binits{M.}},
\oauthor{\bsnm{Park}, \binits{J.}},
\oauthor{\bsnm{Mishra}, \binits{P.}},
\oauthor{\bsnm{Rideout}, \binits{J.R.}},
\oauthor{\bsnm{Sharma}, \binits{D.}},
\oauthor{\bsnm{Dubey}, \binits{S.}},
\oauthor{\bsnm{Inchaurrandieta}, \binits{M.}},
\oauthor{\bsnm{Reiter}, \binits{G.}},
\oauthor{\bsnm{Goel}, \binits{D.}},
\oauthor{\bsnm{Dacie}, \binits{S.}},
\oauthor{\bsnm{Jacob}},
\oauthor{\bsnm{Cetusic}, \binits{G.}},
\oauthor{\bsnm{Taylor}, \binits{G.}},
\oauthor{\bsnm{Meszaros}, \binits{T.}},
\oauthor{\bsnm{Bray}, \binits{E.M.}},
\oauthor{\bsnm{Eigenbrot}, \binits{A.}},
\oauthor{\bsnm{Zahniy}, \binits{S.}},
\oauthor{\bsnm{Zivadinovic}, \binits{L.}},
\oauthor{\bsnm{Parkhi}, \binits{U.}},
\oauthor{\bsnm{Robitaille}, \binits{T.}},
\oauthor{\bsnm{J}, \binits{A.}},
\oauthor{\bsnm{Chicrala}, \binits{A.}},
\oauthor{\bsnm{Ankit}},
\oauthor{\bsnm{Guennou}, \binits{C.}},
\oauthor{\bsnm{D'Avella}, \binits{D.}},
\oauthor{\bsnm{Williams}, \binits{D.}},
\oauthor{\bsnm{Ballew}, \binits{J.}},
\oauthor{\bsnm{Murphy}, \binits{N.}},
\oauthor{\bsnm{Lodha}, \binits{P.}},
\oauthor{\bsnm{Surve}, \binits{R.}},
\oauthor{\bsnm{Bose}, \binits{A.}},
\oauthor{\bsnm{Augspurger}, \binits{T.}},
\oauthor{\bsnm{Krishan}, \binits{Y.}},
\oauthor{\bsnm{neerajkulk}},
\oauthor{\bsnm{Habib}, \binits{I.}},
\oauthor{\bsnm{Letts}, \binits{J.}},
\oauthor{\bsnm{Kothari}, \binits{Y.}},
\oauthor{\bsnm{Keşkek}, \binits{D.}},
\oauthor{\bsnm{honey}},
\oauthor{\bsnm{Molina}, \binits{C.}},
\oauthor{\bsnm{Streicher}, \binits{O.}},
\oauthor{\bsnm{Gomillion}, \binits{R.}},
\oauthor{\bsnm{Wiedemann}, \binits{B.M.}},
\oauthor{\bsnm{Mampaey}, \binits{B.}},
\oauthor{\bsnm{Hill}, \binits{A.}},
\oauthor{\bsnm{Stern}, \binits{K.A.}},
\oauthor{\bsnm{Mittal}, \binits{G.}},
\oauthor{\bsnm{Verstringe}, \binits{F.}},
\oauthor{\bsnm{Dover}, \binits{F.M.}},
\oauthor{\bsnm{Arias}, \binits{E.}},
\oauthor{\bsnm{Stone}, \binits{B.}},
\oauthor{\bsnm{Kannojia}, \binits{S.}},
\oauthor{\bsnm{Kustov}, \binits{A.}},
\oauthor{\bsnm{Yadav}, \binits{T.}},
\oauthor{\bsnm{Wilkinson}, \binits{T.D.}},
\oauthor{\bsnm{Pereira}, \binits{T.M.D.}},
\oauthor{\bsnm{mridulpandey}},
\oauthor{\bsnm{Smith}, \binits{A.}},
\oauthor{\bsnm{Dang}, \binits{T.K.}},
\oauthor{\bsnm{Mehrotra}, \binits{A.}},
\oauthor{\bsnm{Price-Whelan}, \binits{A.}},
\oauthor{\bsnm{B}, \binits{A.}},
\oauthor{\bsnm{yasintoda}},
\oauthor{\bsnm{Stevens}, \binits{A.L.}},
\oauthor{\bsnm{Agrawal}, \binits{Y.}},
\oauthor{\bsnm{Gyenge}, \binits{N.}},
\oauthor{\bsnm{Schoentgen}, \binits{M.}},
\oauthor{\bsnm{abijith-bahuleyan}},
\oauthor{\bsnm{Mendero}, \binits{M.}},
\oauthor{\bsnm{Mangaonkar}, \binits{M.}},
\oauthor{\bsnm{Cheung}, \binits{M.}},
\oauthor{\bsnm{Mekala}, \binits{R.R.}},
\oauthor{\bsnm{Hiware}, \binits{K.}},
\oauthor{\bsnm{Mishra}, \binits{R.}},
\oauthor{\bsnm{Krishna}, \binits{K.}},
\oauthor{\bsnm{Buitrago-Casas}, \binits{J.C.}},
\oauthor{\bsnm{Shashank}, \binits{S.}},
\oauthor{\bsnm{Wimbish}, \binits{J.}},
\oauthor{\bsnm{Calixto}, \binits{J.}},
\oauthor{\bsnm{Babuschkin}, \binits{I.}},
\oauthor{\bsnm{Mathur}, \binits{H.}},
\oauthor{\bsnm{Srikanth}, \binits{S.}},
\oauthor{\bsnm{jamescalixto}},
\oauthor{\bsnm{Kumar}, \binits{G.}},
\oauthor{\bsnm{Gyenge}, \binits{N.G.}},
\oauthor{\bsnm{Murray}, \binits{S.A.}}:
2021,
Sunpy.
\doiurl{10.5281/zenodo.4579839}.
\url{https://doi.org/10.5281/zenodo.4579839}.
\end{botherref}
\endbibitem

\bibitem[\protect\citeauthoryear{Perley and Butler}{2017}]{Perley_2017}
\begin{barticle}
\bauthor{\bsnm{Perley}, \binits{R.A.}},
\bauthor{\bsnm{Butler}, \binits{B.J.}}:
\byear{2017},
\batitle{An accurate flux density scale from 50 mhz to 50 ghz}.
\bjtitle{The Astrophysical Journal Supplement Series}
\bvolume{230}(\bissue{1}),
\bfpage{7}.
\doiurl{10.3847/1538-4365/aa6df9}.
\burl{https://dx.doi.org/10.3847/1538-4365/aa6df9}.
\end{barticle}
\endbibitem

\bibitem[\protect\citeauthoryear{{Pesnell}, {Thompson}, and
  {Chamberlin}}{2012}]{pesnell2012}
\begin{barticle}
\bauthor{\bsnm{{Pesnell}}, \binits{W.D.}},
\bauthor{\bsnm{{Thompson}}, \binits{B.J.}},
\bauthor{\bsnm{{Chamberlin}}, \binits{P.C.}}:
\byear{2012},
\batitle{{The Solar Dynamics Observatory (SDO)}}.
\bjtitle{\solphys}
\bvolume{275}(\bissue{1-2}),
\bfpage{3}.
\doiurl{10.1007/s11207-011-9841-3}.
\adsurl{https://ui.adsabs.harvard.edu/abs/2012SoPh..275....3P}.
\end{barticle}
\endbibitem

\bibitem[\protect\citeauthoryear{{Ramesh}, {Kathiravan}, and
  {Narayanan}}{2011}]{ramesh2011}
\begin{barticle}
\bauthor{\bsnm{{Ramesh}}, \binits{R.}},
\bauthor{\bsnm{{Kathiravan}}, \binits{C.}},
\bauthor{\bsnm{{Narayanan}}, \binits{A.S.}}:
\byear{2011},
\batitle{{Low-frequency Observations of Polarized Emission from Long-lived
  Non-thermal Radio Sources in the Solar Corona}}.
\bjtitle{\apj}
\bvolume{734}(\bissue{1}),
\bfpage{39}.
\doiurl{10.1088/0004-637X/734/1/39}.
\adsurl{https://ui.adsabs.harvard.edu/abs/2011ApJ...734...39R}.
\end{barticle}
\endbibitem

\bibitem[\protect\citeauthoryear{{Sakurai}}{1971}]{sakurai1971}
\begin{barticle}
\bauthor{\bsnm{{Sakurai}}, \binits{K.}}:
\byear{1971},
\batitle{{Energetic Electrons Associated with Solar Flares and Their Relation
  to Type I Noise Activity}}.
\bjtitle{\solphys}
\bvolume{16}(\bissue{1}),
\bfpage{198}.
\doiurl{10.1007/BF00154512}.
\adsurl{https://ui.adsabs.harvard.edu/abs/1971SoPh...16..198S}.
\end{barticle}
\endbibitem

\bibitem[\protect\citeauthoryear{{Subramanian} and
  {Cairns}}{2011}]{subramanian2011}
\begin{barticle}
\bauthor{\bsnm{{Subramanian}}, \binits{P.}},
\bauthor{\bsnm{{Cairns}}, \binits{I.}}:
\byear{2011},
\batitle{{Constraints on coronal turbulence models from source sizes of noise
  storms at 327 MHz}}.
\bjtitle{Journal of Geophysical Research (Space Physics)}
\bvolume{116}(\bissue{A3}),
\bfpage{A03104}.
\doiurl{10.1029/2010JA015864}.
\adsurl{https://ui.adsabs.harvard.edu/abs/2011JGRA..116.3104S}.
\end{barticle}
\endbibitem

\bibitem[\protect\citeauthoryear{Swarup}{2000}]{Swarup2000}
\begin{bbook}
\bauthor{\bsnm{Swarup}, \binits{G.}}:
\byear{2000},
\bbtitle{The Giant Metrewave Radio Telescope},
\bpublisher{American Geophysical Union (AGU)}, \blocation{???},
\bfpage{297}.
\bisbn{9781118668368}.
\doiurl{https://doi.org/10.1029/GM119p0297}.
\burl{https://agupubs.onlinelibrary.wiley.com/doi/abs/10.1029/GM119p0297}.
\end{bbook}
\endbibitem

\bibitem[\protect\citeauthoryear{{The CASA Team} {et~al.}}{2022}]{CASA2022}
\begin{barticle}
\bauthor{\bsnm{{The CASA Team}}},
\bauthor{\bsnm{Bean}, \binits{B.}},
\bauthor{\bsnm{Bhatnagar}, \binits{S.}},
\bauthor{\bsnm{Castro}, \binits{S.}},
\bauthor{\bsnm{Meyer}, \binits{J.D.}},
\bauthor{\bsnm{Emonts}, \binits{B.}},
\bauthor{\bsnm{Garcia}, \binits{E.}},
\bauthor{\bsnm{Garwood}, \binits{R.}},
\bauthor{\bsnm{Golap}, \binits{K.}},
\bauthor{\bsnm{Villalba}, \binits{J.G.}},
\bauthor{\bsnm{Harris}, \binits{P.}},
\bauthor{\bsnm{Hayashi}, \binits{Y.}},
\bauthor{\bsnm{Hoskins}, \binits{J.}},
\bauthor{\bsnm{Hsieh}, \binits{M.}},
\bauthor{\bsnm{Jagannathan}, \binits{P.}},
\bauthor{\bsnm{Kawasaki}, \binits{W.}},
\bauthor{\bsnm{Keimpema}, \binits{A.}},
\bauthor{\bsnm{Kettenis}, \binits{M.}},
\bauthor{\bsnm{Lopez}, \binits{J.}},
\bauthor{\bsnm{Marvil}, \binits{J.}},
\bauthor{\bsnm{Masters}, \binits{J.}},
\bauthor{\bsnm{McNichols}, \binits{A.}},
\bauthor{\bsnm{Mehringer}, \binits{D.}},
\bauthor{\bsnm{Miel}, \binits{R.}},
\bauthor{\bsnm{Moellenbrock}, \binits{G.}},
\bauthor{\bsnm{Montesino}, \binits{F.}},
\bauthor{\bsnm{Nakazato}, \binits{T.}},
\bauthor{\bsnm{Ott}, \binits{J.}},
\bauthor{\bsnm{Petry}, \binits{D.}},
\bauthor{\bsnm{Pokorny}, \binits{M.}},
\bauthor{\bsnm{Raba}, \binits{R.}},
\bauthor{\bsnm{Rau}, \binits{U.}},
\bauthor{\bsnm{Schiebel}, \binits{D.}},
\bauthor{\bsnm{Schweighart}, \binits{N.}},
\bauthor{\bsnm{Sekhar}, \binits{S.}},
\bauthor{\bsnm{Shimada}, \binits{K.}},
\bauthor{\bsnm{Small}, \binits{D.}},
\bauthor{\bsnm{Steeb}, \binits{J.-W.}},
\bauthor{\bsnm{Sugimoto}, \binits{K.}},
\bauthor{\bsnm{Suoranta}, \binits{V.}},
\bauthor{\bsnm{Tsutsumi}, \binits{T.}},
\bauthor{\bparticle{van} \bsnm{Bemmel}, \binits{I.M.}},
\bauthor{\bsnm{Verkouter}, \binits{M.}},
\bauthor{\bsnm{Wells}, \binits{A.}},
\bauthor{\bsnm{Xiong}, \binits{W.}},
\bauthor{\bsnm{Szomoru}, \binits{A.}},
\bauthor{\bsnm{Griffith}, \binits{M.}},
\bauthor{\bsnm{Glendenning}, \binits{B.}},
\bauthor{\bsnm{Kern}, \binits{J.}}:
\byear{2022},
\batitle{Casa, the common astronomy software applications for radio astronomy}.
\bjtitle{Publications of the Astronomical Society of the Pacific}
\bvolume{134}(\bissue{1041}),
\bfpage{114501}.
\doiurl{10.1088/1538-3873/ac9642}.
\burl{https://dx.doi.org/10.1088/1538-3873/ac9642}.
\end{barticle}
\endbibitem

\bibitem[\protect\citeauthoryear{{The SunPy Community}
  {et~al.}}{2020}]{sunpy_community2020}
\begin{barticle}
\bauthor{\bsnm{{The SunPy Community}}},
\bauthor{\bsnm{Barnes}, \binits{W.T.}},
\bauthor{\bsnm{Bobra}, \binits{M.G.}},
\bauthor{\bsnm{Christe}, \binits{S.D.}},
\bauthor{\bsnm{Freij}, \binits{N.}},
\bauthor{\bsnm{Hayes}, \binits{L.A.}},
\bauthor{\bsnm{Ireland}, \binits{J.}},
\bauthor{\bsnm{Mumford}, \binits{S.}},
\bauthor{\bsnm{Perez-Suarez}, \binits{D.}},
\bauthor{\bsnm{Ryan}, \binits{D.F.}},
\bauthor{\bsnm{Shih}, \binits{A.Y.}},
\bauthor{\bsnm{Chanda}, \binits{P.}},
\bauthor{\bsnm{Glogowski}, \binits{K.}},
\bauthor{\bsnm{Hewett}, \binits{R.}},
\bauthor{\bsnm{Hughitt}, \binits{V.K.}},
\bauthor{\bsnm{Hill}, \binits{A.}},
\bauthor{\bsnm{Hiware}, \binits{K.}},
\bauthor{\bsnm{Inglis}, \binits{A.}},
\bauthor{\bsnm{Kirk}, \binits{M.S.F.}},
\bauthor{\bsnm{Konge}, \binits{S.}},
\bauthor{\bsnm{Mason}, \binits{J.P.}},
\bauthor{\bsnm{Maloney}, \binits{S.A.}},
\bauthor{\bsnm{Murray}, \binits{S.A.}},
\bauthor{\bsnm{Panda}, \binits{A.}},
\bauthor{\bsnm{Park}, \binits{J.}},
\bauthor{\bsnm{Pereira}, \binits{T.M.D.}},
\bauthor{\bsnm{Reardon}, \binits{K.}},
\bauthor{\bsnm{Savage}, \binits{S.}},
\bauthor{\bsnm{Sipőcz}, \binits{B.M.}},
\bauthor{\bsnm{Stansby}, \binits{D.}},
\bauthor{\bsnm{Jain}, \binits{Y.}},
\bauthor{\bsnm{Taylor}, \binits{G.}},
\bauthor{\bsnm{Yadav}, \binits{T.}},
\bauthor{\bsnm{Rajul}},
\bauthor{\bsnm{Dang}, \binits{T.K.}}:
\byear{2020},
\batitle{The sunpy project: Open source development and status of the version
  1.0 core package}.
\bjtitle{The Astrophysical Journal}
\bvolume{890}.
\doiurl{10.3847/1538-4357/ab4f7a}.
\burl{https://iopscience.iop.org/article/10.3847/1538-4357/ab4f7a}.
\end{barticle}
\endbibitem

\bibitem[\protect\citeauthoryear{{Thompson}, {Moran}, and
  {Swenson}}{2017}]{thompson2017}
\begin{bbook}
\bauthor{\bsnm{{Thompson}}, \binits{A.R.}},
\bauthor{\bsnm{{Moran}}, \binits{J.M.}},
\bauthor{\bsnm{{Swenson}}, \binits{G.W.} \bsuffix{Jr.}}:
\byear{2017},
\bbtitle{{Interferometry and Synthesis in Radio Astronomy, 3rd Edition}}.
\doiurl{10.1007/978-3-319-44431-4}.
\adsurl{https://ui.adsabs.harvard.edu/abs/2017isra.book.....T}.
\end{bbook}
\endbibitem

\bibitem[\protect\citeauthoryear{Van~Rossum and Drake}{2009}]{python3}
\begin{bbook}
\bauthor{\bsnm{Van~Rossum}, \binits{G.}},
\bauthor{\bsnm{Drake}, \binits{F.L.}}:
\byear{2009},
\bbtitle{Python 3 reference manual},
\bpublisher{CreateSpace},
\blocation{Scotts Valley, CA}.
\bisbn{1441412697}.
\end{bbook}
\endbibitem

\bibitem[\protect\citeauthoryear{Virtanen {et~al.}}{2020}]{Scipy2020}
\begin{barticle}
\bauthor{\bsnm{Virtanen}, \binits{P.}},
\bauthor{\bsnm{Gommers}, \binits{R.}},
\bauthor{\bsnm{Oliphant}, \binits{T.E.}},
\bauthor{\bsnm{Haberland}, \binits{M.}},
\bauthor{\bsnm{Reddy}, \binits{T.}},
\bauthor{\bsnm{Cournapeau}, \binits{D.}},
\bauthor{\bsnm{Burovski}, \binits{E.}},
\bauthor{\bsnm{Peterson}, \binits{P.}},
\bauthor{\bsnm{Weckesser}, \binits{W.}},
\bauthor{\bsnm{Bright}, \binits{J.}},
\bauthor{\bsnm{{van der Walt}}, \binits{S.J.}},
\bauthor{\bsnm{Brett}, \binits{M.}},
\bauthor{\bsnm{Wilson}, \binits{J.}},
\bauthor{\bsnm{Millman}, \binits{K.J.}},
\bauthor{\bsnm{Mayorov}, \binits{N.}},
\bauthor{\bsnm{Nelson}, \binits{A.R.J.}},
\bauthor{\bsnm{Jones}, \binits{E.}},
\bauthor{\bsnm{Kern}, \binits{R.}},
\bauthor{\bsnm{Larson}, \binits{E.}},
\bauthor{\bsnm{Carey}, \binits{C.J.}},
\bauthor{\bsnm{Polat}, \binits{{\. I}.}},
\bauthor{\bsnm{Feng}, \binits{Y.}},
\bauthor{\bsnm{Moore}, \binits{E.W.}},
\bauthor{\bsnm{{VanderPlas}}, \binits{J.}},
\bauthor{\bsnm{Laxalde}, \binits{D.}},
\bauthor{\bsnm{Perktold}, \binits{J.}},
\bauthor{\bsnm{Cimrman}, \binits{R.}},
\bauthor{\bsnm{Henriksen}, \binits{I.}},
\bauthor{\bsnm{Quintero}, \binits{E.A.}},
\bauthor{\bsnm{Harris}, \binits{C.R.}},
\bauthor{\bsnm{Archibald}, \binits{A.M.}},
\bauthor{\bsnm{Ribeiro}, \binits{A.H.}},
\bauthor{\bsnm{Pedregosa}, \binits{F.}},
\bauthor{\bsnm{{van Mulbregt}}, \binits{P.}},
\bauthor{\bsnm{{SciPy 1.0 Contributors}}}:
\byear{2020},
\batitle{{{SciPy} 1.0: Fundamental Algorithms for Scientific Computing in
  Python}}.
\bjtitle{Nature Methods}
\bvolume{17},
\bfpage{261}.
\doiurl{10.1038/s41592-019-0686-2}.
\adsurl{https://rdcu.be/b08Wh}.
\end{barticle}
\endbibitem

\bibitem[\protect\citeauthoryear{{Zhang}, {Wang}, and
  {Kontar}}{2021}]{zhang2021parametric}
\begin{barticle}
\bauthor{\bsnm{{Zhang}}, \binits{P.}},
\bauthor{\bsnm{{Wang}}, \binits{C.}},
\bauthor{\bsnm{{Kontar}}, \binits{E.P.}}:
\byear{2021},
\batitle{{Parametric Simulation Studies on the Wave Propagation of Solar Radio
  Emission: The Source Size, Duration, and Position}}.
\bjtitle{\apj}
\bvolume{909}(\bissue{2}),
\bfpage{195}.
\doiurl{10.3847/1538-4357/abd8c5}.
\adsurl{https://ui.adsabs.harvard.edu/abs/2021ApJ...909..195Z}.
\end{barticle}
\endbibitem

\bibitem[\protect\citeauthoryear{{Zlobec}}{1971}]{zlobec1971}
\begin{bchapter}
\bauthor{\bsnm{{Zlobec}}, \binits{P.}}:
\byear{1971},
\bctitle{{Polarization of type I bursts}}.
In: \beditor{\bsnm{{Abrami}}, \binits{A.}} (ed.)
\bbtitle{CESRA-2, Committee of European Solar Radio Astronomers}
\bseriesno{2},
\bfpage{101}.
\adsurl{https://ui.adsabs.harvard.edu/abs/1971cesra...2..101Z}.
\end{bchapter}
\endbibitem

\bibitem[\protect\citeauthoryear{{Zlobec} {et~al.}}{1992}]{zlobec1992}
\begin{barticle}
\bauthor{\bsnm{{Zlobec}}, \binits{P.}},
\bauthor{\bsnm{{Messerotti}}, \binits{M.}},
\bauthor{\bsnm{{Dulk}}, \binits{G.A.}},
\bauthor{\bsnm{{Kucera}}, \binits{T.}}:
\byear{1992},
\batitle{{VLA and Trieste Observations of Type-I Storms - Type-Iv and
  Pulsations}}.
\bjtitle{\solphys}
\bvolume{141}(\bissue{1}),
\bfpage{165}.
\doiurl{10.1007/BF00155910}.
\adsurl{https://ui.adsabs.harvard.edu/abs/1992SoPh..141..165Z}.
\end{barticle}
\endbibitem

\end{thebibliography}

\end{article} 

\end{document}